\documentclass[twocolumn,showpacs,preprintnumbers,amsmath,amssymb,pre,,superscriptaddress]{revtex4-1}

\usepackage{xcolor}    
\usepackage{graphicx}
\usepackage{dcolumn}
\usepackage{bm}
\usepackage{subfigure}
\usepackage{amssymb}
\usepackage{multirow}
\usepackage{mathtools}
\usepackage{braket}
\usepackage{placeins}

\graphicspath{{plots/}}

\usepackage{braket}
\usepackage{upgreek}

\usepackage{tikz}
\usepackage{upgreek}
\usetikzlibrary{shapes,backgrounds,calc,patterns}
\tikzset{subfiglabel/.style={left, minimum size=4ex}}%

\usepackage{ncmds}

\newcommand{\gud}{g^{\uparrow\downarrow}}
\newcommand{\guu}{g^{\uparrow\uparrow}}

\newcommand{\rpdm}{d}
\newcommand{\trpdm}{d}

\renewcommand{\pi}{\uppi}

\renewcommand{\crt}{\hat{a}^{\dagger}} 
\renewcommand{\ani}{\hat{a}}
\renewcommand{\cre}[1]{\crt_{#1}}
\renewcommand{\an}[1]{\ani_{#1}}
\renewcommand{\n}{\hat{n}} 
\renewcommand{\fcrt}{\hat{\Psi}^{\dagger}}
\renewcommand{\fani}{\hat{\Psi}}
\renewcommand{\fcre}[1]{\fcrt_{#1}}
\renewcommand{\fan}[1]{\fani_{#1}}

\renewcommand{\vec}[1]{\mathbf{#1}}

\renewcommand{\vr}{\vec{r}}
\newcommand{\vk}{\vec{k}}
\newcommand{\occconfig}[1]{\{#1\}}

\begin{document}

\title{Momentum distribution function and short-range correlations of the warm dense electron gas -- ab initio quantum Monte Carlo results}

\author{Kai Hunger}
\affiliation{
 Institut f\"ur Theoretische Physik und Astrophysik, Christian-Albrechts-Universit\"at zu Kiel,
 Leibnizstra{\ss}e 15, 24098 Kiel, Germany}

 \author{Tim Schoof}
\affiliation{
 Institut f\"ur Theoretische Physik und Astrophysik, Christian-Albrechts-Universit\"at zu Kiel,
 Leibnizstra{\ss}e 15, 24098 Kiel, Germany}
 
 \affiliation{
 Deutsches Elektronen Synchotron (DESY), Hamburg, Germany}

\author{Tobias Dornheim}
\affiliation{Center for Advanced Systems Understanding (CASUS), D-02826 G\"orlitz, Germany}
\affiliation{Helmholtz-Zentrum Dresden-Rossendorf (HZDR), D-01328 Dresden, Germany}

\author{Michael Bonitz}
\affiliation{
 Institut f\"ur Theoretische Physik und Astrophysik, Christian-Albrechts-Universit\"at zu Kiel,
 Leibnizstra{\ss}e 15, 24098 Kiel, Germany}

\author{Alexey Filinov}
\affiliation{
 Institut f\"ur Theoretische Physik und Astrophysik, Christian-Albrechts-Universit\"at zu Kiel,
 Leibnizstra{\ss}e 15, 24098 Kiel, Germany}
 \affiliation{
 Joint Institute for High Temperatures, Russian Academy of Sciences, Izhorskaya 13, Moscow 125412, Russia}

\begin{abstract}
In a classical plasma the momentum distribution, $n(k)$,  decays exponentially, for large $k$, and the same is observed for an ideal Fermi gas. However, when quantum and correlation effects are relevant simultaneously, an algebraic decay, $n_\infty(k)\sim k^{-8}$ has been predicted. This is of relevance for cross sections and threshold processes in dense plasmas that depend on the number of energetic particles.
Here we present extensive \textit{ab initio} results for the momentum distribution of the nonideal uniform electron gas at warm dense matter conditions. Our results are based on first principle fermionic path integral  Monte Carlo (CPIMC) simulations and clearly confirm the $k^{-8}$ asymptotic. This asymptotic behavior is directly linked to short-range correlations which are analyzed  via the on-top pair distribution function (on-top PDF), i.e. the PDF of electrons with opposite spin. We present extensive results for the density and temperature dependence of the on-top PDF and for the momentum distribution in the entire momentum range.
\end{abstract}

\pacs{xxx}

\maketitle
\section{Introduction}
  Dense quantum plasmas and warm dense matter (WDM) are attracting growing interest in recent years. Typical for WDM are densities around solid densities and elevated temperatures around the Fermi temperature, e.g. \cite{graziani-book,dornheim_physrep_18,Fortov2016,bonitz_pop_20}. Such situations are common in astrophysical systems \cite{chabrier_quantum_1993,schlanges-etal.95cpp,bezkrovny_pre_4,nettelmann_saturn_2013}, including the interiors of giant planets  and white dwarf stars, or the atmosphere of neutron stars. In the laboratory, WDM situations are realized upon laser or ion beam compression of matter \cite{Ernstorfer1033} and also in experiments on inertial confinement fusion (ICF) \cite{hurricane_inertially_2016, hu_militzer_PhysRevLett.104.235003}. Under WDM conditions the electrons are typically quantum degenerate and moderately correlated whereas ions are classical and, possibly strongly correlated. These properties clearly manifest themselves in the thermodynamic \cite{filinov_ppcf_01,militzer_massive_2008,Militzer_2013,Militzer_PRE_2021, dornheim_prl16},
 transport and optical properties~\cite{witte_prl_17,PhysRevE.71.016409,PhysRevE.73.036401, hamann_prb_20,dornheim_prl_18,hamann_cpp_20} of WDM. To gain deeper understanding of this unusual state of matter, accurate results for structural quantities are essential, including the pair distribution function \cite{FILINOV_pla_00,militzer_path_2000} and the static~\cite{dornheim_cpp17} and dynamic structure factor~\cite{dornheim_prl_18,groth_prb_19,Kraus_2018,redmer_glenzer_2009}. For additional investigations of the  uniform electron gas model at finite temperature, see Refs.~\cite{ksdt,PhysRevB.99.195134,PhysRevB.88.115123,PhysRevB.62.16536,dornheim_physrep_18}. 
 
 Here we consider another many-particle property -- the momentum distribution function $n(k)$ and how it is influenced by finite temperature and Coulomb interaction effects.
It is well known that, for classical systems in thermodynamic equilibrium,  $n(k)$ is always of Maxwellian form regardless of the strength of the interaction. In contrast, in a quantum system the momentum and coordinate dependencies do not decouple which leads to fundamentally different behaviors of $n(k)$ in ideal and nonideal quantum systems, and only for an ideal system the familiar Fermi distribution, $n^{\rm id}(k)$ is being recovered (here we consider only Fermi systems). However, in a non-ideal Fermi system, the momentum distribution decays much slower with $k$, exhibiting a power law asymptotic.   The importance of a power law asymptotic has been pointed out by Starostin and co-workers \cite{starostin_quantum_2002,starostin_ppr05,starostin_jetp17} and many others, e.g. \cite{savchenko_pop01},  because an increased number of particles in high-momentum states could have a significant effect on scattering and reaction cross sections, in particular on fusion reaction rates in dense plasmas \cite{salpeter_69,ichimaru_RevModPhys.65.255,dewitt_ctpp.2150390124}. The main goal of the present paper is, therefore, to present accurate theoretical results for the tail of the momentum distribution function. Before outlining our goals in more detail, we briefly recall the main available theoretical results on the large-$k$ asymptotic of the momentum distribution function.

It was first demonstrated by Wigner~\cite{wigner_quantum_1932} how to incorporate quantum uncertainty between coordinate and momentum into $n(k)$. 
Following the development of perturbation theory for the electron gas in the 1950's, e.g.~\cite{bohm_collective_1953},~\cite{nozieres_correlation_1958}, Daniel and Vosko~\cite{daniel_vosko_momentum_1960}  calculated the momentum distribution for an interacting electron gas. They used the approximation due to Gell-Mann and Brueckner for the correlation energy~\cite{gell1957correlation} which corresponds to the random rhase Approximation (RPA). 
For the ground state, $T=0$ K, they derived an analytical expression for the large-$k$ asymptotic of the momentum distribution,
\begin{equation}
    \lim_{k\to \infty} n^{\mathrm{RPA}}(k) 
    \sim \frac{1}{{k^8}}\,,
    \label{eq:k8}
\end{equation}
i.e. they found an algebraic decay, in striking contrast to the exponential asymptotic of an ideal classical or quantum system.

Galitskii and Yakimets~\cite{galitskii_particle_1967} used Matsubara Green functions and the Kadanoff-Baym relation \cite{kadanoff-baym} between the energy distribution in equilibrium, $f^{\rm EQ}(\omega)$ [which is always a Fermi or Bose distribution], and the spectral function $A(k,\omega)$, 
\begin{align}
     n(k) = \int \frac{d\omega}{2\pi} A(k,\omega) f^{\rm EQ}(\omega)\,.
     \label{eq:kba}
\end{align}
Correlation effects enter only via the spectral function $A$, which is given by $A^{\rm id}(k,\omega)=2\pi\delta[\hbar \omega - E(k)]$, for an ideal gas. Ref.~\cite{galitskii_particle_1967} computed the leading correction to the ideal spectral function and confirmed the asymptotic, Eq.~(\ref{eq:k8}).
For a systematic improvement of this result higher order selfenergies have been computed, e.g. by Kraeft \textit{et al}. Ref.~\cite{kraeft_pre_02}, and we also refer to the text books Refs.~\cite{kadanoff-baym,bonitz_qkt,balzer-book}.

The exact limiting behavior in the asymptotic (\ref{eq:k8}) was found
independently by Kimball~\cite{kimball_short_range_1975} 
via a short-range ansatz to the two-electron wave function,  and by Yasuhara and Kawazoe~\cite{yasuhara_note_1976} who analyzed the large-momentum behavior of the ladder terms in Goldstone perturbation theory. 
An important result of Yasuhara \textit{et al.} is the proof \cite{yasuhara_note_1976} that, at $T=0$ K, the asymptotic can be expressed via the on-top pair distribution function (on-top PDF), i.e. the PDF of a particle pair with different spin projections at zero distance, $g^{\uparrow\downarrow}(r=0)$,
\begin{align}
    \lim_{k\to \infty} n(k)  = \frac{4}{9}\left(\frac{4}{9\pi}\right)^{2/3}\left( \frac{r_s}{\pi}\right)^2 \frac{k^8_F}{k^8}g^{\uparrow\downarrow}(0)\,,
      \label{eq:k8-g0}
\end{align}
where $k_F$ denotes the Fermi momentum, and the coupling (Brueckner) parameter $r_s=\bar r/a_B$ is the ratio of the mean interparticle distance, $\bar r = [3/(4\pi n)]^{1/3}$, to the Bohr radius~\cite{Ott2018}.
A more general derivation has been presented by Hofmann \textit{et al.}~\cite{hofmann_short-distance_2013} who have shown that Eq.~(\ref{eq:k8-g0}) holds also for finite temperature.

An extension of the results of Yasuhara \textit{et al.} and Kimball to arbitrary spin polarizations of the electron gas was performed by Rajagopal \textit{et al.} in Ref.~\cite{rajagopal_short_ranged_1978} who derived the next order in the asymptotic which becomes dominant in the case of a ferromagnetic electron gas because the on-top PDF vanishes:
\begin{equation}
    \label{eq:g0_k10}
    n^{\rm ferro}(k) \xrightarrow[{k\to \infty}]{}
    \frac{4}{3} \frac{8}{9\pi^2} \frac{g^{\uparrow\uparrow ''}(0)}{2} \frac{(\alpha \rs)^2}{k^{10}}
    .
\end{equation}
Aside from dense plasmas, the tail of the momentum distribution is also relevant for the electron gas in metals, e.g. \cite{holzmann_prl11}, as well as cold fermionic atoms \cite{jensen_prl20,doggen_momentum_resolved_2015}. In the latter case, however, the short-range character of the pair interaction leads to a modified large-momentum asymptotic, $n(k) \sim k^{-4}$, instead of (\ref{eq:k8}).

A second approach to the high-momentum tail is based on quantum Monte Carlo simulations. Here one can either directly compute the asymptotic of $n(k)$ or determine it from the Fourier transform of the density matrix. While the former requires to extend the simulations to very large momenta and to resolve the occupations over many orders of magnitude, the latter way is  potentially more efficient.  
Here one calculates the on-top PDF
(which is called ``contact'' in the cold atomic gas community). 
In addition to its use in Eq.~(\ref{eq:k8-g0}), we mention that an accurate description of $g^{\uparrow\downarrow}(0)$ is interesting in its own right, and is important for many other applications, like the description of the static local field correction~\cite{dornheim2020effective,holas_limit,Sjostrom_Gradient_2014,dornheim_ML,Takada_PRB_2016}.

Accurate QMC results for $n(k)$ of the UEG in the ground state were obtained in Refs.~\cite{PhysRevB.44.7879,holzmann_prl11}, whereas the on-top PDF was studied in multiple QMC-based works~\cite{ortiz_prb_94,holzmann_prl11,PhysRevLett.82.5317,PhysRevB.61.7353}, most recently by Spink and co-workers~\cite{Spink_Drummond_PRB_2013}. At finite temperatures, the momentum distribution $n(k)$ has been investigated by Militzer \textit{et al.} \cite{Militzer_PRL_2002,Militzer_HEDP_2019} who carried out restricted path integral Monte Carlo (RPIMC) simulations and recently by Filinov \textit{et al.} \cite{larkin_cpp18} based on a version of fermionic PIMC that is formulated in phase space. 
Furthermore, the only comprehensive data set for $g(0)$ in this regime was presented by Brown \textit{et al.}~\cite{Brown_2014}, again on the basis of RPIMC simulations.

Note that fermionic PIMC in coordinate space is limited to moderate degeneracy \cite{filinov_ppcf_01,filinov_pre15}, due to the notorious fermion sign problem, see Ref.~\cite{dornheim_pre_2019} for an accessible topical discussion. On the other hand, 
RPIMC has been shown to exhibit significant systematic errors of the thermodynamic quantities, for example the error for the exchange-correlation energy reaches $10\%$ at $r_s=1$ and $\Theta=0.25$ \cite{schoof_prl15}.
In addition, RPIMC is substantially hampered by an additional sampling problem (\emph{reference point freezing}~\cite{Brown_chapter}) at high densities, $r_s\lesssim1$.

Therefore, it is of high interest to perform alternative simulations that can access the momentum distribution of the uniform electron gas at high degeneracy without any systematic errors.
In this context, a suitable approach is given by the recently developed configuration PIMC (CPIMC) method
that is formulated in Fock space (Slater determinant space) and is highly efficient at high to moderate quantum degeneracy \cite{schoof_cpp15, schoof_prl15}.
In particular, CPIMC simulations were the basis for 
the first \textit{ab initio} thermodynamic results for the warm dense UEG \cite{schoof_prl15}. In combination with the likewise novel permutation blocking PIMC \cite{dornheim_njp15, dornheim_jcp15,Dornheim_CPP_2019} scheme, it was possible to avoid the fermion sign problem and to obtain \textit{ab initio} thermodynamic results for the UEG at warm dense matter conditions
\cite{groth_prl17, dornheim_physrep_18}. 
In addition, also \textit{ab initio} results for the static density response \cite{groth_jcp17} have been obtained with CPIMC.

The goal of this paper is to utilize CPIMC to obtain \textit{ab initio} data for the momentum distribution  of the  uniform electron gas at finite temperature and high density corresponding to $r_s \lesssim 0.7$. To access stronger coupling, we also employ a recently developed approximate method -- restricted CPIMC \cite{yilmaz_jcp_20} as well as direct fermionic propagator PIMC simulations in coordinate space -- an extension of permutation blocking PIMC \cite{dornheim_njp15}.
 In particular, 
\begin{description}
\item[i] we verify that the high-momentum asymptotic does obey a $k^{-8}$ behavior, and that it is solely determined by the on-top PDF;
\item[ii] we present detailed CPIMC results for $g^{\uparrow\downarrow}(0)$ and analyze its temperature and density dependence;
\item[iii] investigate the momentum distribution function in the vicinity of the Fermi momentum and for small momenta;
\item[iv] investigate the momentum range of the onset of the large-momentum asymptotic.
\end{description} 

This paper is organized as follows: In Sec.~\ref{s:theory-overvies} we present a brief overview on earlier theoretical work pertaining to the uniform electron gas, together with the main predictions. This is followed by an introduction into our quantum Monte Carlo simulations in Sec.~\ref{s:qmc_wdeg} and by a presentation of the numerical results in Sec.~\ref{s:results}.


 \section{Theory framework }\label{s:theory-overvies}
\subsection{On-top pair distribution}\label{ss:otpdf}
Since the high-momentum tail of the momentum distribution function can be expressed in terms of the on-top pair distributions, cf. Eq.~(\ref{eq:k8-g0}), we start by considering the pair distribution of electrons with spin projections $\sigma_1$ and $\sigma_2$~\cite{giuliani2005quantum},
\begin{equation}
    \label{eq:pair_distribution_spinresolved_def}
    \begin{aligned}
    g_{\sigma_1\sigma_2}(\vec{r}_1,\vec{r}_2) &=
    \frac{ \Braket{
        \fcre{\sigma_1}(\vec{r}_1) \fcre{\sigma_2}(\vec{r}_2) \fan{\sigma_2}(\vec{r}_2) \fan{\sigma_1}(\vec{r}_1)
    } }{
        \Braket{\fcre{\sigma_1}(\vec{r}_1) \fan{\sigma_1}(\vec{r}_1)}
        \Braket{\fcre{\sigma_2}(\vec{r}_2) \fan{\sigma_2}(\vec{r}_2)}
    }\,,
    \end{aligned}
\end{equation}
where $\Psi_{\sigma_1}(\textbf{r}_1)$ [$\Psi^\dagger_{\sigma_1}(\textbf{r}_1)$] is a fermionic field operator annihilating [creating] an electron in spin state $|\textbf{r}_1 \sigma_1\rangle$.
Note that the two-particle density in the numerator is normalized to the single-particle spin densities, $n_{\sigma}(\textbf{r})=\langle\fcre{\sigma}(\vec{r}) \fan{\sigma}(\vec{r})\rangle$, in the denominator. Thus, in the absence of correlations and exchange effects, $g_{\sigma_1\sigma_2}(\vec{r}_1,\vec{r}_2) \equiv 1$.
For electrons there exist four spin combinations. Assuming a homogeneous paramagnetic system, we have $g^{\uparrow\uparrow}(\textbf{r}_1,\textbf{r}_2)\equiv g^{\downarrow\downarrow}(\textbf{r}_1,\textbf{r}_2)$ and $g^{\uparrow\downarrow}(\textbf{r}_1,\textbf{r}_2)\equiv g^{\downarrow\uparrow}(\textbf{r}_1,\textbf{r}_2)$.

The total pair distribution function follows from the spin-resolved functions (\ref{eq:pair_distribution_spinresolved_def}) according to
\begin{align}
    \label{eq:pair_distribution_def1}
    g(\vec{r}_1,\vec{r}_2) &=
    \sum\limits_{\sigma_1 \sigma_2} g_{\sigma_1\sigma_2}(\vec{r}_1,\vec{r}_2)\frac{n_{\sigma_1}(\textbf{r}_1)n_{\sigma_2}(\textbf{r}_2)}{n(\textbf{r}_1)n(\textbf{r}_2)}\,,
\\
n(\textbf{r})&= \sum_{\sigma} n_{\sigma}(\textbf{r})\,,
\label{eq:density}
\end{align}
where the normalization assures that, in the absence of exchange and correlation effects, $g\equiv 1$.
In a spatially homogeneous system, such as the UEG, the PDFs depend only on the distance of the pair, \(g_{\sigma_1\sigma_2}(\vec{r}_1,\vec{r}_2)= g_{\sigma_1\sigma_2}(|\vec{r}_2 - \vec{r}_1|)\). Of particular importance is the case of zero separation. Then, the Pauli principle leads to 
$g^{\uparrow\uparrow}(0)\equiv g^{\downarrow\downarrow}(0)\equiv 0$. On the other hand, the probability of
finding two electrons with different spins ``on top of each other'' yields the on-top PDF, $\gud(0)$, which is related to total PDF in the paramagnetic case by [cf. Eq.~(\ref{eq:pair_distribution_def1})]
\begin{align}
g(0) = \frac{\guu(0) + \gud(0)}{2} = \frac{1}{2}\gud(0)\,,  
\label{eq:on-top-pdf}
\end{align}
which is a fundamental property for the characterization of short-range correlations. While in a non-interacting  system ($r_s \to 0$), $\gud_{\rm id}(0)=1$, Coulomb repulsion leads to a reduction of this value. Thus for the UEG a monotonic reduction with $r_s$ is expected which will directly influence, via Eq.~(\ref{eq:k8-g0}), the tail of the momentum distribution.

There exist a variety of analytical parametrizations of the on-top PDF.
The ground state on-top PDF of correlated electrons was investigated in Ref.~\cite{gori_giorgi_short_range_2001} by using the Overhauser screened Coulomb potential in the radial two-particle Schr\"odinger equation.
The results were parametrized for \(\rs\leq 10\) according to
\begin{equation}
    \label{eq:g0_overhausermodel}
    \gud(0) = (1.0 + A \rs + B \rs^2 + C \rs^3 + D \rs^4) \mathrm{e}^{-E \rs}\,,
\end{equation}
where \(A=0.0207\), \(B=0.08193\), \(C=-0.01277\), \(D=0.001859\) and \(E=0.7524\). These results will be called ``Overhauser model'' and used for comparison below.

On the other hand, the  high-temperature asymptotic of the on-top PDF of a classical non-degenerate electron gas where $\chi = n\Lambda^3 \ll 1$, and $\Theta=k_BT/E_F \gg 1$, is also known. Here $n$ is the density depending on the mean inter-particle distance, $\bar r \sim n^{-1/3}$, and $\Lambda$ is the thermal DeBroglie wavelength, $\Lambda^2=h^2/(2\pi mk_BT)$. A quantum-mechanical expansion was given in Ref.~\cite{hofmann_short-distance_2013}, where the result depends on the order the high-temperature limit, $T\to \infty$, and the classical limit, $\hbar \to 0$, are taken. The reason is the existence of a third length scale \cite{green-book,bonitz_qkt}, the Bjerrum length, $l_B=\beta e^2$, where $\beta = (k_{\mathrm{B}}T)^{-1}$, giving rise to a second dimensionless parameter, the classical coupling parameter, $\Gamma=\beta e^2/\bar r = l_B/\bar r$.

In the case $\Gamma \ll \chi^{1/3}$ (i.e. $l_B \ll \Lambda$), the result is \cite{hofmann_short-distance_2013} %
\begin{equation}
    \label{eq:pair_corr_hightemperature_expansion_barth}
    g(0) = \frac{1}{2}\left(1 - \sqrt{2}\pi\,\frac{l_B}{\Lambda} + \ldots \right)\,,
\end{equation}
where the behavior is still dominated by the ideal Fermi gas properties with deviations  scaling like $\Gamma \chi^{-1/3}$, or, $(k_BT)^{-1/2}n^0$.

On the other hand, in the case $\chi^{1/3} \ll \Gamma$ (i.e., $\Lambda \ll l_B$), which corresponds to classical plasmas at moderate temperatures, the on-top PDF becomes \cite{hofmann_short-distance_2013}
\begin{equation}
    \label{eq:pair_corr_hightemperature_expansion_barth}
    g(0) =  \frac{4\pi^2 2^{1/3}}{3^{1/3}}\left(\frac{l_B}{\Lambda}\right)^{4/3} e^{-\frac{3\pi}{2^{1/3}}\left(\frac{l_B}{\Lambda}\right)^{2/3}}+ \ldots \,.
\end{equation}
This value is exponentially small due to the moderate Coulomb repulsion and is not influenced by quantum effects. Nevertheless, quantum effects (finite $\Lambda$) show up in the algebraic momentum tail, according to Eq.~(\ref{eq:k8-g0}), but only on length scales much smaller than $\Lambda$ or, correspondingly, at momenta strongly exceeding $\Lambda^{-1}$.
The latter case is out of the range of WDM and not relevant for the present analysis.

Finally, there exists a more recent parametrization of the ground state on-top-PDF that is based on QMC simulations~\cite{Spink_Drummond_PRB_2013}: 
\begin{align}
g(0; r_s)= \frac{1+a \sqrt{r_s}+b r_s}{1+c r_s+ d r_s^3}\,, \quad T=0\,\mbox{K}\,,
\label{eq:g0-spink}    
\end{align}
which will be used for comparison below.
For an overview about different models of $g(0)$ for the ground state, the reader is referred to the paper by Takada~\cite{Takada_PRB_2016}.
With explicit results for the on-top PDF and, using Eq.~(\ref{eq:k8-g0}), the large-$k$ asymptotics of the momentum distribution function can be reconstructed.

For finite temperature one can relate the PDF to an effective quantum pair potential, $g^{\uparrow\downarrow}(r) =e^{-\beta V_Q(r)}$, an idea that was put forward by Kelbg \cite{kelbg_ap_63_1} and further developed, among others, by Deutsch, Ebeling, and Filinov and co-workers, cf.~Refs.~\cite{deutsch_pla_77,filinov_jpa03, filinov_pre04,ebeling_jpa_06} and references therein. We will return to this issue in Sec.~\ref{sss:t-dependence}.

\subsection{Configuration PIMC (CPIMC) approach to $g(0)$ and $n(k)$ of the warm dense electron gas}\label{s:qmc_wdeg}

\subsubsection{Idea of CPIMC simulations }\label{ss:cpimc}
CPIMC was first formulated in Ref.~\cite{schoof_cpp11} and applied to the UEG in Refs.~\cite{schoof_cpp15,schoof_prl15,groth_prb16}.
For a detailed description of the CPIMC formalism we refer to the overview articles~\cite{cpimc_springer_14,dornheim_physrep_18} and to the recent developments \cite{yilmaz_jcp_20}.
Here we only summarize the main idea. The thermodynamic expectation value of an arbitrary operator $\hat A$ is determined by the density operator $\hat \rho$ and its normalization -- the partition function $Z$, where we use the canonical ensemble,
\begin{align}
    \hat \rho &= e^{-\beta \hat H}\,,
    \label{eq:rho-def}
\quad    Z(\beta) = \mbox{Tr}\, \hat \rho\,,
\\
    \langle \hat A\rangle(\beta) &=\frac{1}{Z}
    \mbox{Tr}\,  \hat A \hat \rho\,.
    \label{eq:mean-a}
\end{align}
Since the Hamiltonian involves only one- and two-body operators, 
\begin{align}
\label{eq:interaction_hamiltonian}
\hat H = \sum_{ij} h_{ij} \hat a^\dagger_i \hat a_j 
 + \frac{1}{2}\sum_{ijkl} w_{ijkl} \, \hat a^\dagger_i  a^\dagger_j\hat a_l\hat a_k\,, 
\end{align}
its expectation value can be described via the reduced one- and two-particle density matrices, \(\rpdm_{ij}\) and \(\trpdm_{ijkl}\), see the definitions (\ref{eq:TDrelation_1pdm}) and (\ref{eq:TDrelation_2pdm}). Here the sums are over arbitrary complete sets of single-particle states which below  will be specified to momentum eigenstates. Quantum Monte Carlo estimators for these quantities are obtained through differentiation of the partition function~\cite[Eq.~(5.88)]{cpimc_springer_14} with respect to the single-particle matrix element
\begin{equation}
    \label{eq:TDrelation_1pdm}
    \rpdm_{ij}
    \coloneqq
    \braket{\cre{i}\an{j}}
    =
    -\frac{1}{\beta} \del{h_{ij}} \ln Z \,,
\end{equation}
and the two-particle matrix element
\begin{equation}
    \label{eq:TDrelation_2pdm}
    d_{ijkl}
    \coloneqq
    \braket{\cre{i}\cre{j}\an{k}\an{l}}
    =
    -\frac{1}{\beta} \del{w_{ijkl}} \ln Z,
\end{equation}
respectively. 
The resulting expressions depend on the order and choice of the indices \((i,j)\) and \((i,j,k,l)\), respectively.

Let us now present explicit expressions for the one-particle and two-particle density matrices in CPIMC. Configuration PIMC is path integral Monte Carlo formulated in Fock space \cite{schoof_cpp11}, i.e. in the space of $N$-particle Slater determinants, $|\{n\}\rangle=|\{n_1,n_2,\dots\}\rangle$, constructed from the single-particle orbitals $|i\rangle$ where $n_i$ is the associated occupation number.

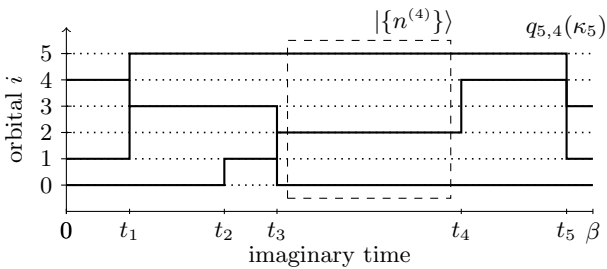
\begin{figure}[ht]
    \centering

\begin{tikzpicture}[xscale=0.7, yscale=0.35]
\newcommand{\xrange}{10}
\newcommand{\yrange}{1}

\newcommand{\taunull}{0}
\newcommand{\taunulllb}{$0$}
\newcommand{\taueins}{\xrange*0.12}
\newcommand{\taueinslb}{$t_1$}
\newcommand{\tauzwei}{\xrange*0.3}
\newcommand{\tauzweilb}{$t_2$}
\newcommand{\taudrei}{\xrange*0.4}
\newcommand{\taudreilb}{$t_3$}
\newcommand{\tauvier}{\xrange*0.75}
\newcommand{\tauvierlb}{$t_4$}
\newcommand{\taufuenf}{\xrange*0.95}
\newcommand{\taufuenflb}{$t_5$}
\newcommand{\taulast}{\xrange}
\newcommand{\taulastlb}{$\beta$}
\newcommand{\zero}{\yrange*1}
\newcommand{\eins}{\yrange*2}
\newcommand{\zwei}{\yrange*3}
\newcommand{\drei}{\yrange*4}
\newcommand{\vier}{\yrange*5}
\newcommand{\fuenf}{\yrange*6}

\draw (0,0) -- +(\xrange,0) coordinate (xlabel);
\draw[->] (0,0) -- +(0,\yrange*7) coordinate (ylabel);
\foreach \i in {0,...,5} {
	\draw (-0.1,\i*\yrange+\yrange) node[left] {$\i$} -- (0.1,\i*\yrange+\yrange);
}
\foreach \i/\l in {\taunull/\taunulllb,\taueins/\taueinslb,\tauzwei/\tauzweilb,\taudrei/\taudreilb,\tauvier/\tauvierlb,\taufuenf/\taufuenflb,\taulast/\taulastlb} {
\draw (\i,0.1) -- (\i,-0.1) node[below] {\l};
}
\draw (0,0.1) -- (0,-0.1) node[below] {$0$};
\node at (0.5*\xrange,-1.7) {imaginary time};
\node[rotate=90] at (-1,3.5*\yrange) {orbital $i$};

\foreach \i in {1,...,6} {
\draw[semithick,dotted] (0,\i*\yrange) -- (\xrange,\i*\yrange);
}

\begin{scope}[thick]
\draw (0,\zero) -| (\tauzwei,\eins) -| (\taudrei,\zero) -- (\xrange,\zero);
\draw (0,\eins) -| (\taueins,\drei) -| (\taudrei,\zwei) -| (\tauvier,\vier) -| (\taufuenf,\drei) -- (\xrange,\drei);
\draw (0,\vier) -| (\taueins,\fuenf) -| (\taufuenf,\eins) -- (\xrange,\eins);
\draw (\taueins, \drei) -- (\taueins, \vier);
\draw (\taudrei, \eins) -- (\taudrei, \zwei);
 \end{scope}

\node at (\taufuenf,\yrange*7) {$q_{5,4}(\kappa_5)$};

\draw[dashed] (\taudrei+0.2,0.5) rectangle (\tauvier-0.2,\yrange*6.5);
\node at (\tauzwei*2.2,\fuenf+\yrange+0.25) {$\ket{\onv{4}}$};

\end{tikzpicture}
    
    \caption{Illustration of a path $C$, Eq.~(\ref{eq:path2}), with five kinks. The three kinks 1, 3, 5, at times $t_1$, $t_3$, $t_5$, each involve four orbitals: $\kappa_1=(1,4;3,5)$, $\kappa_3=(1,3;0,2)$, $\kappa_5=(4,5;1,3)$ respectively. The two kinks 2 and 4, at $t_2$ and $t_4$, involve two orbitals, each: $\kappa_2=(0;1)$ and $\kappa_4=(2;4)$. The fourth Slater determinant $|n^{(4)}\rangle$ exists between the imaginary ``times'' $t_3$ and $t_4$ and contains three occupied orbitals \(\set{0,2,5}\).}
    \label{fig:Kink_path}
\end{figure}

In CPIMC the canonical partition function~(\ref{eq:rho-def}) is written as a Dyson series in imaginary time, for details see Ref.~\cite{yilmaz_jcp_20}. A \emph{configuration} \(C\) determining a MC state is given by a set of initially occupied orbitals \(\onvv\), along with a set of \(K\) \emph{changes} \(\kappa_i\) to this set, called \emph{kinks} at their respective times \(t_i\), \(1\leq i\leq K\),
\begin{equation}
    C \coloneqq \Set{\onvv, t_1,\ldots,t_K, \kappa_1,\ldots,\kappa_K}\,.
    \label{eq:path2}
\end{equation}
Due to the Slater-Condon rules for fermionic 2-particle operators, each interaction matrix element yields either a 2-particle term, corresponding to \(\kappa = (i,j)\), or a 4-particle term, \(\kappa = (i,j,k,l)\).  Thus the kinks are given by either two or four orbital indices, respectively. The \emph{kink matrix element} \(\qi{i}(\kappa_i)\) represent the off-diagonal matrix elements with respect to the possible choices of 2- or 4-tuples \(\kappa_i\). The final result for the partition function is \cite{yilmaz_jcp_20}
\begin{widetext}
\begin{equation}
    \label{eq:partition_perturbation_compact_kinks}
    Z(\beta)
    =
    \sum_
{\substack{K=0 \\ K \neq 1}}^{\infty}
    \sum\limits_{\onvv}
    \sum\limits_{\kappa_1} \ldots \sum\limits_{\kappa_K}
    \int\limits_0^{\beta} \dx t_1 
    \int\limits_{t_1}^{\beta} \dx t_2 \ldots 
    \int\limits_{t_{K-1}}^{\beta} \dx t_K
    \;
    (-1)^K
    \left(
    \prod\limits_{i=0}^{K} 
    \euler^{
    -
    \Di{i}(t_{i+1} - t_i)
    }
    \right)
    \times
    \left(
    \prod\limits_{i=1}^{K} 
    \qi{i}(\kappa_i)
    \right)\,,
\end{equation}
\end{widetext}
where paths with $K=1$ violate the periodicity and have to be excluded. Configurations can be sampled from the partition function
\begin{equation}
    Z = \intsum_C W(C)\,,
\end{equation}
with the weight function
\begin{equation}
    W(C) 
    =
    (-1)^K
    \left(
    \prod\limits_{i=0}^{K} 
    \euler^{
    -
    \Di{i}(t_{i+1} - t_i)
    }
    \right)
    \left(
    \prod\limits_{i=1}^{K} 
    \Yij{i}{i-1}
    \right)\,,
\label{eq:weight-c}
\end{equation}
which allows one to rewrite thermodynamic expectation values (\ref{eq:mean-a}) as 
\begin{equation}
    \braket{A} = \intsum_C W(C) A(C)\,  .
\end{equation}
An example configuration (path) is illustrated in Fig.~\ref{fig:Kink_path}. With three particles present, horizontal solid lines represent diagonal matrix elements, as given by the exponential factor in the partition function (\ref{eq:partition_perturbation_compact_kinks}) and the occupation number state at a given time-interval is specified by the set of all these lines in this interval. On the other hand, the vertical solid lines represent interaction terms, where the occupation changes according to the specified \emph{kink} \(\kappa_i\), weighted by the respective kink matrix element \(\qi{i}(\kappa_i)\). Due to the periodicity of the expectation values~(\ref{eq:mean-a}), the kinks must add to yield the initial occupation vector at \(0<t<t_1\) again:
\begin{gather}
    \prod\limits_{i=1}^{K} \hat{q}_{i,i-1}(\kappa_i)
    =
    \hat{\mathbf{1}}
    .
\end{gather}

This representation of the partition function can now be applied to the observables of interest.
For the one-particle density matrix we obtain, for \(i\neq j\), 
\begin{equation}
    \label{eq:estimator_1pdm_offdiag}
    d_{ij}(C) = -\fro{\beta} \sum_{\nu=1}^K 
    \frac{(-1)^{\alpha_{\occconfig{n^{(\nu)}},i,j}}}{q_{\occconfig{n^{(\nu)}}\occconfig{n^{(\nu-1)}}}(\kappa_\nu)} \delta_{\kappa_\nu,(i,j)}
    .
\end{equation}
For the uniform electron gas, the off-diagonal matrix elements vanish in a momentum basis, whereas the diagonal ones yield the momentum distribution, as will be discussed in Sec.~\ref{ss:cpimc-nk}

Let us now turn to the CPIMC estimator for the two-particle density matrix. Here we have to distinguish several cases of index combinations~\cite[Eq.~3.14]{schoof_cpimc_2016}.
If \(i<j,k<l\) are pairwise distinct
\begin{equation}
    \label{eq:estimator_2pdm_offdiag}
    d_{ijkl}(C) = -\fro{\beta}\sum_{\nu=1}^K 
    \frac{(-1)^{\alpha_{\occconfig{n^{(\nu)}},i,j} + \alpha_{\occconfig{n^{(\nu-1)}},k,l}}}{q_{\occconfig{n^{(\nu)}}\occconfig{n^{(\nu-1)}}}(\kappa_\nu)} \delta_{\kappa_\nu,(i,j,k,l)}.
\end{equation}
The term under the sum (without the Kronecker-delta) will be abbreviated as the \emph{weight} of the kink \(\kappa_\nu\),
\begin{equation*}
    \mathcal{W}(\kappa_{\nu}) \coloneqq \frac{(-1)^{\alpha_{\occconfig{n^{(\nu)}},i,j} + \alpha_{\occconfig{n^{(\nu-1)}},k,l}}}{q_{\occconfig{n^{(\nu)}}\occconfig{n^{(\nu-1)}}}(\kappa_\nu)}\,.
\end{equation*}
In the case of \(i=k\), but with all other indices being different,
\begin{align}
	d_{ijil}(C) = -\fro{\beta}\sum_{\nu=1}^K \frac{(-1)^{\alpha_{\occconfig{n^{(\nu)}},j,l}}}{q_{\occconfig{n^{(\nu)}}\occconfig{n^{(\nu-1)}}}(\kappa_\nu)} n_i^{(\nu)} \delta_{\kappa_\nu,(j,l)}\,.
	\label{eq:estimator_2pdm_halfdiag}
\end{align}
Finally, if \(i=k\) and \(j=l\), but \(i\neq j\), the matrix elements are given by
\begin{align}
	d_{ijij}(C) = \sum_{\nu=0}^{K} n_i^{(\nu)}  n_j^{(\nu)} \frac{\tau_{\nu+1}-\tau_\nu}{\beta}.
	\label{eq:estimator_2pdm_diag}
\end{align}
The expectation value of this estimator is given by the weighted sum over all possible configurations C,
\begin{equation}
    \label{eq:expval_2pdm}
    d_{ijkl} = \Braket{ \cre{i}\cre{j}\an{k}\an{l} } = \fro{Z} \intsum\limits_C  d_{ijkl}(C) W(C)\,.
\end{equation}

Due to the large single-particle basis sizes that have to be used in the CPIMC simulations, the variances of these estimators may be very large for some transitions [i.e. combinations of indices (i,j) or (i,j,k,l)]. However, special cases can be used to derive the estimators needed to measure short-range properties of the system: The momentum distribution and the on-top PDF.

\subsubsection{Momentum distribution with CPIMC}\label{ss:cpimc-nk}
The momentum distribution function is given by the diagonal part of  Eq.~(\ref{eq:TDrelation_1pdm}), if a plane wave basis is being used, cf.~Sec.~\ref{ss:cpimc-g0}. For \(i=j\), we obtain 
\begin{equation}
    \label{eq:CPIMC_estimator_nk}
    \begin{aligned}
    \braket{\n_i} 
    &= -\frac{1}{\beta} \del{h_{ii}} \log(Z)
= \frac{1}{Z} \intsum\limits_{C}\;
    n_i(C)  W(C)
    \\
n_i(C) &=     \sum\limits_{\nu=0}^K
    n_i^{(\nu)}
    \frac{ \tau_{i+1} - \tau_i}{\beta}\,,
    \end{aligned}
\end{equation}
where the contribution of each time-slice is weighted by the length of horizontal paths.

\subsubsection{On-top pair distribution function with CPIMC}\label{ss:cpimc-g0}
The definition~(\ref{eq:pair_distribution_spinresolved_def}) of the spin-resolved PDF requires the two-particle density matrix in coordinate representation which is obtained from the two-particle density matrix, Eq.~(\ref{eq:TDrelation_2pdm}), in momentum representation, i.e. using plane wave orbitals,
\begin{equation}
    \label{eq:plane_waves_single_particle_basis_def}
    \braket{\vr \sigma|\vk s} = 
    \fro{\sqrt{V}} \euler^{\ii\vk\vr} \delta_{s,\sigma} \eqqcolon \varphi_{\vk}(\vr)\delta_{s,\sigma}\,.
\end{equation}
To shorten the notation, the wave vector \(\vk\) will be represented by an index \(i \leftrightarrow \vk_i\) of the corresponding single-particle basis eigenvalue.
%
The field operators in a position-spin basis are related to the creation and annihilation operators in a momentum-spin basis \(\ket{i} \coloneqq \ket{\vk_i s_i}\) by~
\begin{equation}
    \label{eq:creation_annihilation_operators_basis_change}
    \begin{aligned}
    \fan{\sigma}(\vr) &= \sum_i \phi_i(\vr,\sigma)\an{i}\,, \\
    \fcre{\sigma}(\vr) &= \sum_i \phi_i^*(\vr,\sigma) \cre{i}\,.
    \end{aligned}
\end{equation}

The \emph{on-top} PDF, Eq.~(\ref{eq:on-top-pdf}), follows from the PDF, Eq.~(\ref{eq:pair_distribution_spinresolved_def}), for different spin projections, \(\sigma_2\neq\sigma_1\),
\begin{equation}
    \label{eq:ontop_pair_density}
    \gud(0)
    \coloneqq
    g_{\sigma_1\sigma_2}(\vr,\vr)
    \eqqcolon
    \gud_0\,.
\end{equation}
With the basis transformation~(\ref{eq:creation_annihilation_operators_basis_change}) of the field operators, a straightforward calculation yields the CPIMC estimator for the on-top PDF (for details see   Appendix~\ref{app:on-top-pdf}), 
\begin{align}
    \gud_0(C)
    &=
    \fro{\beta}\sum_{\nu=1}^K 
    \sum\limits_{\substack{k\neq i<j\neq l\\k<l}}
 (1 - \delta_{s_{i{\nu}},s_{j{\nu}}}) w(\kappa_{\nu} ) \times
    \nonumber\\
    &  \qquad  \left(
    \delta_{s_{j_{\nu}},s_{l_{\nu}}}\delta_{s_{i_{\nu}},s_{k_{\nu}}}
    - 
    \delta_{s_{i_{\nu}},s_{l_{\nu}}}\delta_{s_{j_{\nu}},s_{k_{\nu}}}
    \right) -
    \nonumber\\
    &
    -
    \sum_{\nu=0}^{K} 
    \sum\limits_{i<j}
    (1 - \delta_{s_i,s_j}) n_i^{(\nu)}  n_j^{(\nu)} \frac{\tau_{\nu+1}-\tau_\nu}{\beta}
    .
    \label{eq:ontop_pairdensity_cpimc_estimator_further2}
\end{align}

\section{Simulation results
}\label{s:results}
%

We have performed extensive CPIMC simulations with $N=54$ particles. Due to the fermion sign problem, these simulations are restricted to small coupling parameters, $r_s\lesssim 0.7$. To extend the range of parameters, we also performed simulations with $N=14$ particles. As shown before,  important structural properties, such as the static structure factor \cite{dornheim_prl16,PhysRevLett.97.076404} and the pair distribution function only weakly depend on the particle number. A quantitative analysis of the $N$-dependence of the results will be performed for the tail of the momentum distribution in Sec.~\ref{ss:g(0)-tdl}. The CPIMC results are complemented by restricted CPIMC simulations \cite{yilmaz_jcp_20}. To access larger values of the coupling parameter, we also include   fermionic PIMC simulation results in coordinate space for the on-top PDF.

\subsection{Momentum distribution}\label{ss:n_p-data}
\subsubsection{Overview}\label{sss:n_p-overview}
\begin{figure}[h!]
    \centering
    \includegraphics[width=\linewidth]{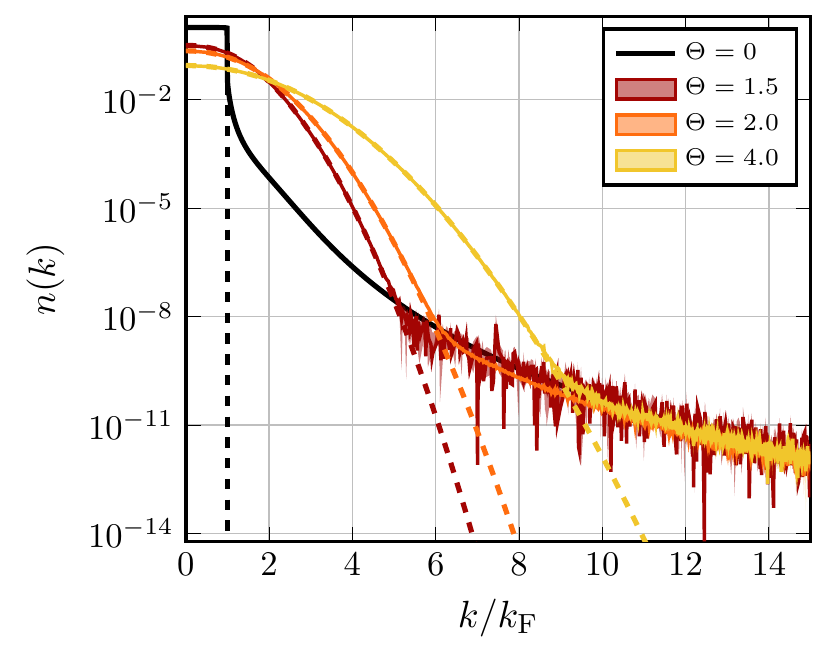}
    \caption{Temperature dependence of the momentum distribution of moderately correlated electrons,   \(\rs=0.5\). CPIMC results with $N=54$ particles for 
    three temperatures are compared to the ground state (solid black, data of Ref.~\cite{gori_giorgi_short_range_2001}). For comparison,  the ideal Fermi distribution is shown by dashed lines of the same color as the interacting result.
}
    \label{fig:fullMDF_N54_rs05}
\end{figure}

Let us start by analyzing the general trends of the momentum distribution when either the temperature or the coupling strength are varied. 
In Fig.~\ref{fig:fullMDF_N54_rs05} we present CPIMC data for $N=54$ particles showing the entire momentum range for moderate coupling, $r_s=0.5$, and three temperatures and indicating that the occupation of high-momentum states is coupled in a non-trivial way to occupation of lower momentum states. Interestingly, an increase of temperature not only leads to the familiar broadening of $n(k)$ around the Fermi edge and depletion below it, but may also lead to a lower population of the tail  (see below). The most striking observation is the strong deviation,  in the tail region, from the exponential decay in case of an ideal Fermi gas. Our simulations clearly confirm the correlation-induced enhanced population of high-momentum states with the asymptotic, $n(k)\sim k^{-8}$.

\begin{figure}[h!]
    \centering
    \includegraphics[width=\linewidth]{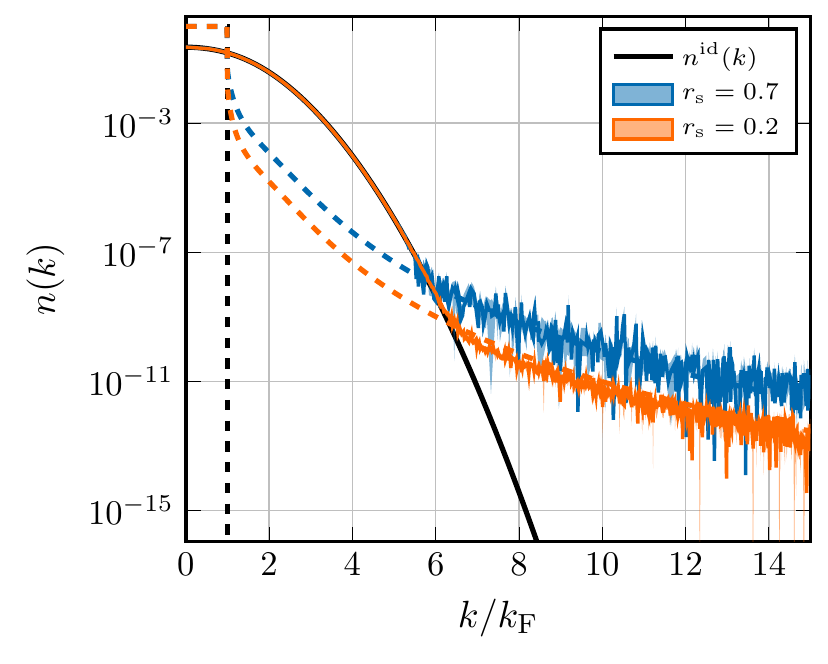}
    \caption{Density dependence of the momentum distribution of moderately correlated electrons at temperature \(\Theta=2\). 
    CPIMC results with $N=54$ particles are compared to the ideal Fermi-Dirac distribution $n^{\rm id}$ (full black line). For momenta below approximately $6k_F$ the correlated distributions are indistinguishable from $n^{\rm id}$.
    For comparison, the ground state distributions, as given by~Ref.~\cite{gori_giorgi_short_range_2001}, are shown by the dashed lines of the same color as the finite temperature result.
    }
    \label{fig:fullMDF_N54_th2}
\end{figure}

Let us now turn to the dependence on the coupling parameter. To this end, we present, in Fig.~\ref{fig:fullMDF_N54_th2}, the momentum distribution for a fixed temperature, $\Theta=2$, and two values of $r_s$ and also compare to the ideal Fermi gas. For large momenta, $k\gtrsim 6 k_F$, we observe an increase of the population when $r_s$ grows. However, for intermediate momenta, $k_F \lesssim k \lesssim 6 k_F$, the ideal distribution is significantly above the correlated distributions. Finally, below the Fermi momentum, the correlated distributions are again above the ideal momentum distribution. 

This behavior seems counter intuitive, and we analyze it more in detail in the next section.
%

\subsubsection{Interaction-induced enhanced population of low-momentum states}\label{sss:low-momentum-states}
\begin{figure}[h!]
    \centering
    \includegraphics[width=\linewidth]{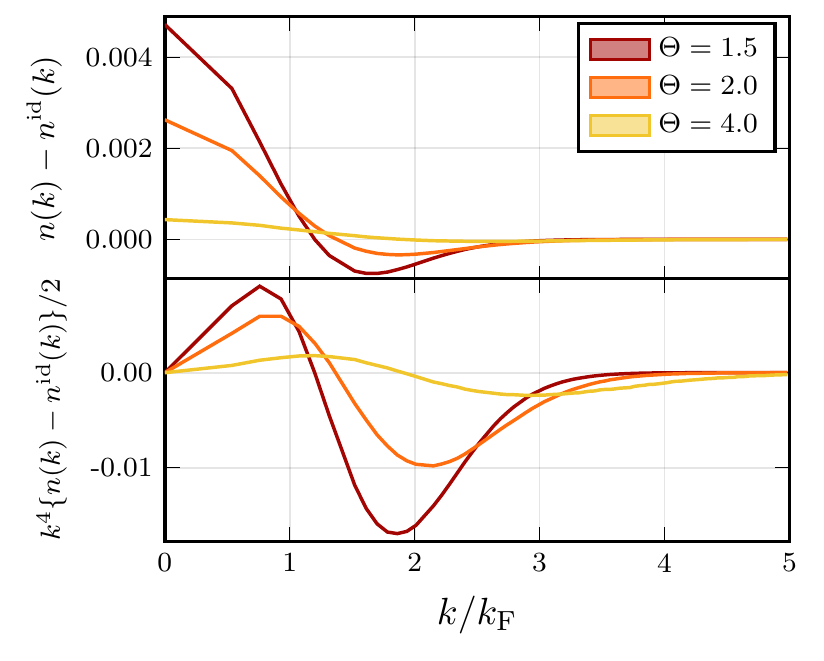}
    \caption{Deviation of the momentum distribution (CPIMC results with $N=54$ particles) from the ideal Fermi-Dirac distribution at moderate coupling \(\rs=0.5\).
    Lower Panel: Difference of the distribution functions weighted with $k^4/2$ (Hartree units), i.e. k-resolved kinetic energy density.
}
    \label{fig:Bulk_N54_rs05}
\end{figure}

\begin{figure}[h!]
    \centering
    \includegraphics[width=\linewidth]{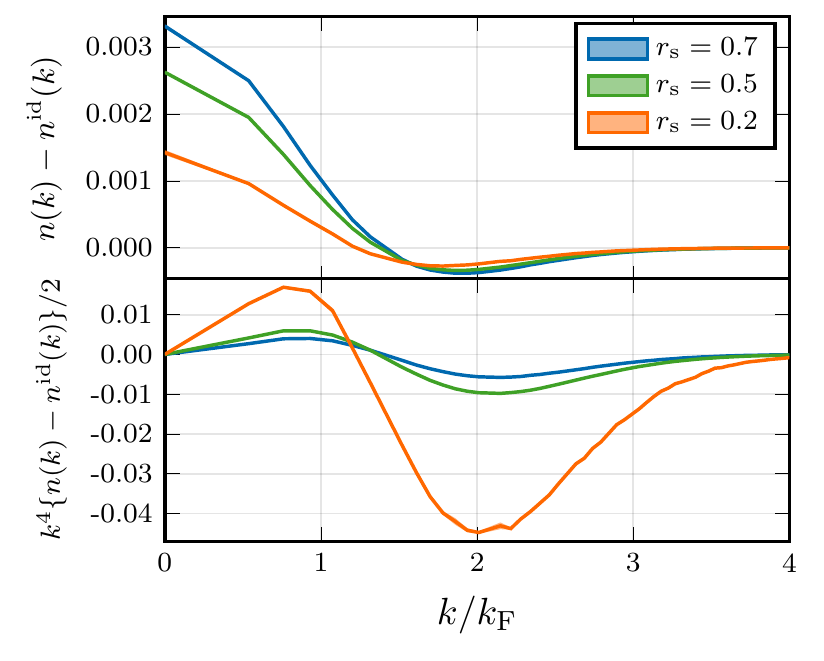}
    \caption{Same as Fig.~\ref{fig:Bulk_N54_rs05}, 
but for a fixed temperature, \(\Theta=2\), and three densities. The different ordering of the curves in the lower panel arises from the $r_s$-dependence of the horizontal scale, $k_F \propto r_s^{-1}$.
    }
    \label{fig:Bulk_N54_th2}
\end{figure}

Let us now investigate in more detail the behavior of the momentum distribution in the range from $k=0$ to momenta on the order of several $k_F$. To focus on correlation effects we plot, in Fig.~\ref{fig:Bulk_N54_rs05}, the difference of the correlated distribution and the Fermi distribution   for the case of $r_s=0.5$. Clearly, we observe an enhanced population of low-momentum states, $k \lesssim 1.5k_F$, compared to the Fermi function. The effect is biggest at the lowest temperature and decreases monotonically with $\Theta$. On the other hand, it is clear that, upon further reduction of $\Theta$, this effect will decrease again and vanish in the ground state. The reason is that, at $T=0$ K, all low-momentum states are completely occupied, and, due to the Paui principle, correlations can only enhance the population of unoccupied states, at $k>k_F$.
The same analysis is performed, for a fixed temperature but different coupling parameters, in Fig.~\ref{fig:Bulk_N54_th2}. Here we observe a monotonic trend: with increasing $r_s$, the difference of the populations increases with respect to the ideal case.

This interaction-induced enhanced population of low-k state has been reported before, e.g. based on restricted PIMC simulations, by Militzer and Pollock \cite{Militzer_PRL_2002}, and on thermodynamic Green functions by Kraeft \textit{et al.} \cite{kraeft_pre_02}. The origin of this effect is interaction-induced lowering of the energy eigenvalues, $E(k) < E^{\rm id}(k)$ \cite{Militzer_PRL_2002}. Here, the interacting energy contains, in addition, an exchange and a correlation contribution, 
\begin{align}
E(k)=E^{\rm id}(k)+\Delta E_{\rm x}(k) + \Delta E_{\rm c}(k)\,.
\label{eq:e-dispersion}
\end{align}
The behavior reported here is dominated by the exchange contribution, i.e. by the Hartree-Fock selfenergy (the Hartree term vanishes due to homogeneity and charge neutrality) which is negative,
\begin{align}
\Delta E_{\rm x}(p)=\Sigma^{\rm HF}(p) &= - \int \frac{d^3q}{(2\pi\hbar )^3} w(|\textbf{p}-\textbf{q}|)\,n(q)\,.
\label{eq:sigma-hf}
\end{align}
The negative Hartree-Fock selfenergy shift is largest at small momenta and decreases monotonically with $k$. As a consequence, the system tends to increase the population of low-momentum states. 

An interesting consequence of this population increase is that the mean kinetic energy of the correlated electron gas may be lower than that of the ideal electron gas at the same temperature \cite{Militzer_PRL_2002,kraeft_pre_02}. Our simulations clearly confirm this prediction. This effect is illustrated in the lower panels of Figs.~\ref{fig:Bulk_N54_rs05} and \ref{fig:Bulk_N54_th2} where we plot the $k$-resolved difference of kinetic energy densities. For the parameters shown in theses figures, the excess kinetic energy (compared to the ideal UEG) concentrated in low-momentum states (positive difference) is smaller than the kinetic energy reduction (negative difference) at larger momenta. This is evident from the areas under the curves in the lower panels of Figs.~\ref{fig:Bulk_N54_rs05} and \ref{fig:Bulk_N54_th2}. As a result the total kinetic energy difference of the interacting system compared to the ideal system is negative for a broad range of parameters. The corresponding kinetic energies for the interacting and ideal systems are presented in the appendix, in tables~\ref{tab:TvsT0_table_N54} and \ref{tab:TvsT0_table_N14}, for 54 and 14 particles, respectively.

Our argument, so far, was based on the negative sign of the Hartree-Fock selfenergy. However, for a complete picture we also need to consider the energy shift due to correlations, $\Delta E_c$.
In contrast to the Hartree-Fock shift, the correlation corrections to the energy dispersion are typically positive, but smaller, as was shown for the Born approximation (Montroll-Ward approximation), in Ref.~\cite{kraeft_pre_02}. However, this result applies only for weak coupling. For stronger coupling, in particular, $r_s \gtrsim 1$, at least T-matrix selfenergies would be required. An alternative are QMC simulations, as presented in Ref.~\cite{Militzer_PRL_2002}, which allow one to map out the range of density and temperature parameters where the difference of correlated and ideal kinetic energies changes sign. 

\begin{figure}
    \centering
    \includegraphics[width=\linewidth]{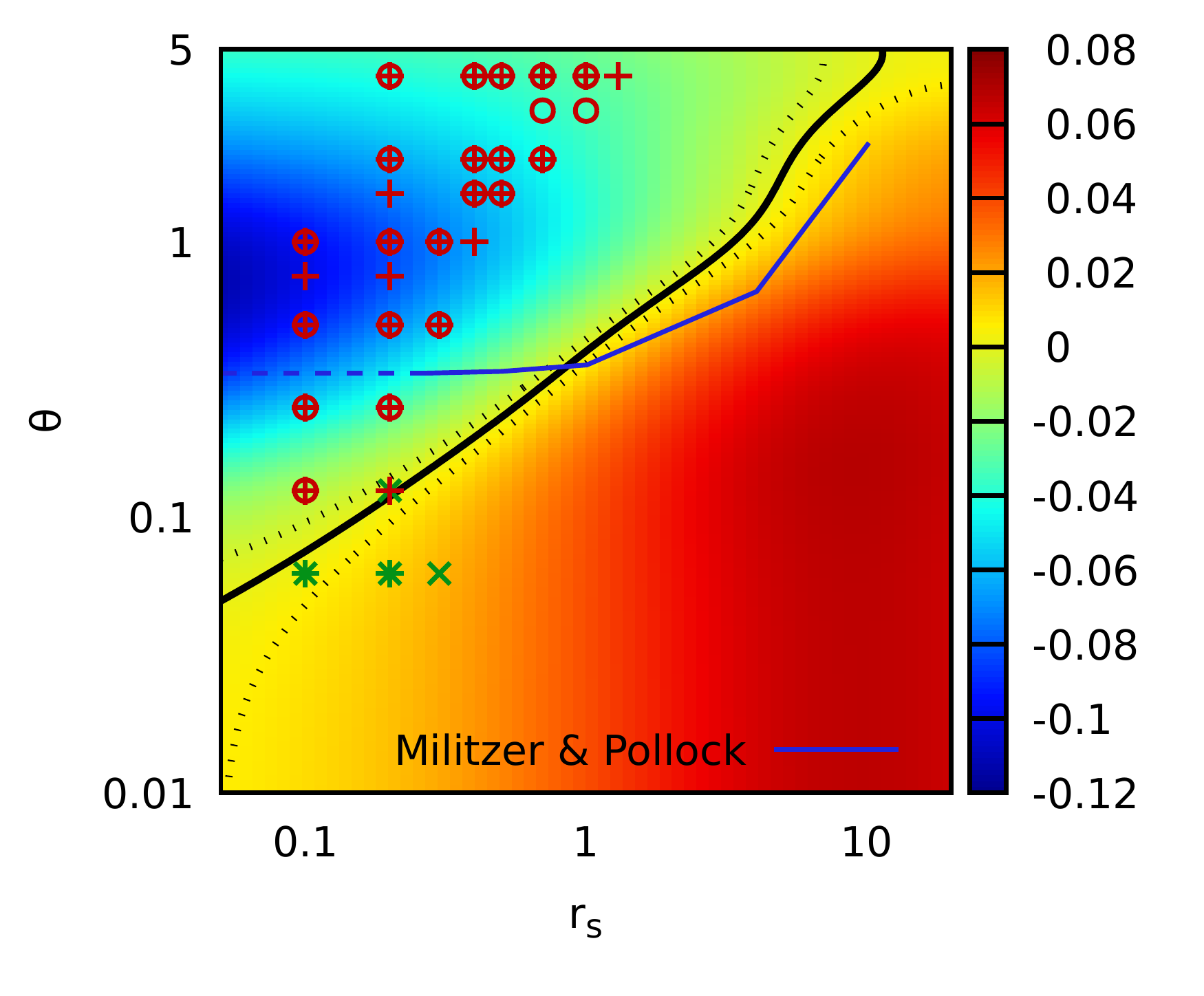}
    \caption{Interaction-induced lowering of the kinetic energy. Heat-map: $K_\textnormal{xc}$, Eq.~(\ref{eq:Kxc}), computed from the parametrization by Groth \textit{et al.}~\cite{groth_prl17}. Solid black line:  $r_s$-$\Theta$-combinations where $K_\textnormal{xc}$ vanishes;  dotted lines: uncertainty interval of $5\times10^{-3}$Ha. Solid blue line:  $K_\textnormal{xc}=0$ according to RPIMC results of Ref.~\cite{Militzer_PRL_2002}.
    Red (green) pluses: CPIMC results for kinetic energy decrease (increase) compared to ideal case. Red circles (green crosses): CPIMC data points where the occupation of the lowest orbital, $n(0)$, is higher (lower) than in the ideal case, i.e. $n^{\rm id}(0)$. Extensive data for the kinetic energy are presented in the tables in the Appendix.
    \label{fig:kxc-overviewl}}
\end{figure}

The present CPIMC simulations are not directly applicable to the range
$r_s \gtrsim 1$. However, we can take advantage of the accurate parametrization of the exchange--correlation free energy $f_\textnormal{xc}$ of Groth \textit{et al. }\cite{groth_prl17} that is based on a combination of CPIMC, PB-PIMC and ground state QMC results.
In particular, the exchange--correlation contribution to the kinetic energy is obtained by evaluating~\cite{groth_prb_19}
 \begin{eqnarray}\label{eq:Kxc}
 K_\textnormal{xc} 
 &=& - f_\textnormal{xc}
 -\theta \frac{\partial f_\textnormal{xc}}{\partial\theta}\Big|_{r_s}
- r_s \frac{\partial f_\textnormal{xc}}{\partial r_s}\Big|_\theta \quad ,
 \end{eqnarray}
and the corresponding results are depicted in Fig.~\ref{fig:kxc-overviewl}.
The line where the kinetic energy difference changes sign is in good agreement with the results of Ref.~\cite{Militzer_PRL_2002}, for $r_s \gtrsim 1$, but we find significant deviations at smaller $r_s$ and lower temperatures. 

It is interesting to compare the parameter values where the kinetic energy difference changes sign to the occupation of the zero-momentum state, $n(0)$, relative to the ideal 
distribution, $n^{\rm id}(0)$. 
For most temperatures considered, the interacting zero-momentum state \(n(0)\) has a larger population than the corresponding ideal state. Only for the lowest temperatures,  \(\Theta\in\set{1/16,1/8}\), we observe the opposite behavior.

\subsubsection{High-momentum asymptotics of $n(k)$}

\begin{figure}[h!]
    \centering
    \includegraphics{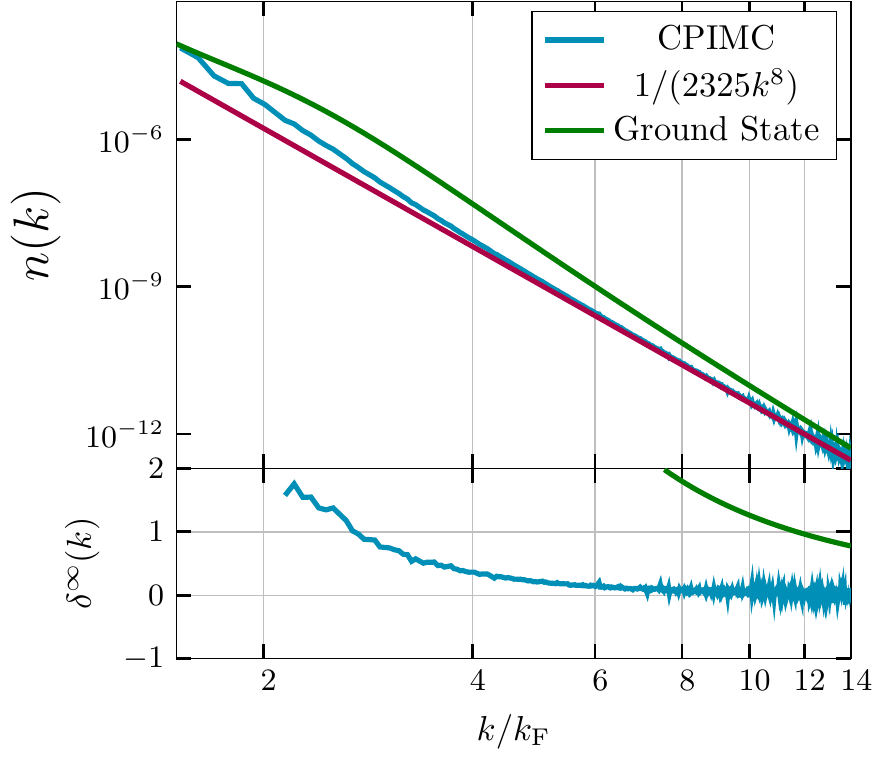}
    \caption{Large momentum behavior of the momentum distribution, for \(\rs=0.2\) and \(\Theta=0.0625\). \textbf{Top}: blue line. CPIMC results for $N=54$ particles, pink line: best fit to the  asymptotic, Eq.~(\ref{eq:k8-g0}), with $g_{\uparrow\downarrow}(0)$ taken from CPIMC data; green: ground state value from~Ref.~\cite{gori-giorgi_momentum_2002}. \textbf{Bottom}: relative difference of CPIMC and the ground state data from the asymptotic (pink line in top plot), according to Eq.~(\ref{eq:relative_difference_tail}). 
    }
    \label{fig:tails_rs02_th00625_N54}
\end{figure}

\begin{figure}[h!]
    \centering
    \includegraphics{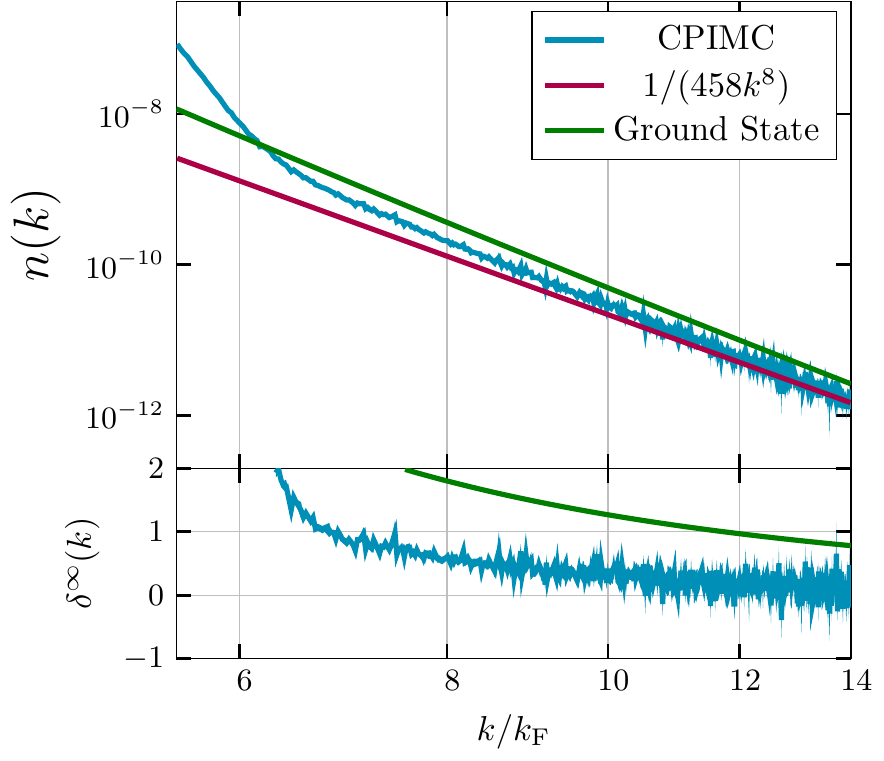}
    \caption{Same as Fig.~\ref{fig:tails_rs02_th00625_N54}, but for $r_s=0.5$ and $\Theta=2.$
}
    \label{fig:tails_rs05_th2_N54}
\end{figure}

In Figs.~\ref{fig:tails_rs02_th00625_N54} and \ref{fig:tails_rs05_th2_N54} we present data for low to moderate temperatures focusing on momenta beyond the Fermi edge. We directly compare the CPIMC data to the asymptopic behavior where a $k^{-8}$ tail is expected, with the coefficient determined by the on-top PDF $g(0)$, cf. Eq.~(\ref{eq:k8-g0}), where $g(0)$ is taken from the same CPIMC simulation. As can be seen in these figures, the CPIMC data clearly exhibit the expected algebraic decay, for sufficiently large $k$. To make a quantitative comparison, we also plot, in the lower panels, the relative difference between CPIMC data, $n(k)$, and the asymptotic, \(n^\infty(k)\), according to \begin{equation}
    \label{eq:relative_difference_tail}
    \delta^\infty(k) = \frac{n(k)}{n^\infty(k)} - 1
    .
\end{equation}
%
%
The results for $\delta^\infty$  clearly confirm that our \textit{ab initio} CPIMC data approach the asymptotic. Moreover, we can estimate the momentum range where the asymptotic behavior dominates. For low temperatures of $E_F/16$, the asymptotic is reached at about $6 k_F$, cf. Fig.~\ref{fig:tails_rs02_th00625_N54}. With increasing temperature, the asymptotic is approached only at larger momenta, e.g. for 
$\Theta=2$, around $11 k_F$, cf. Fig.~\ref{fig:tails_rs05_th2_N54}. A systematic analysis of the onset of the asymptotic will be given in Sec.~\ref{ss:onset}.

In these figures we also included ground state data for the momentum distribution (green lines) which allows us to analyze finite temperature effects. In all figures we observe that the finite temperature distribution, $n(k;\Theta)$, intersects the ground state function, $n(k;0)$, coming from above, before it reaches the asymptotic. In the range of the algebraic tail the finite temperature function is always below the ground state result, for the same $r_s$ and $k$, in agreement with
Fig.~\ref{fig:fullMDF_N54_rs05}. This behavior is, at first sight, counter intuitive because one expects that finite temperature effects increase the population of high momentum states. As we will show in Sec.~\ref{sss:t-dependence} this temperature dependence is, in fact, non-monotonic and is due to a competition between Coulomb repulsion and exchange effects. 

Finally, we note that our simulations reveal that the $k^{-8}$ asymptotic is observed independently of the particle number, in agreement with the predictions of Refs.~\cite{hofmann_short-distance_2013,barth_fewbody_2015}. We will return to the question of the particle number dependence in Sec.~\ref{ss:n-dependenc}.

\subsection{Ab initio results for $g(0)$ 
}\label{ss:g(0)-tdl}

\begin{figure}[h!]
    \centering
    \includegraphics[width=0.5\textwidth]{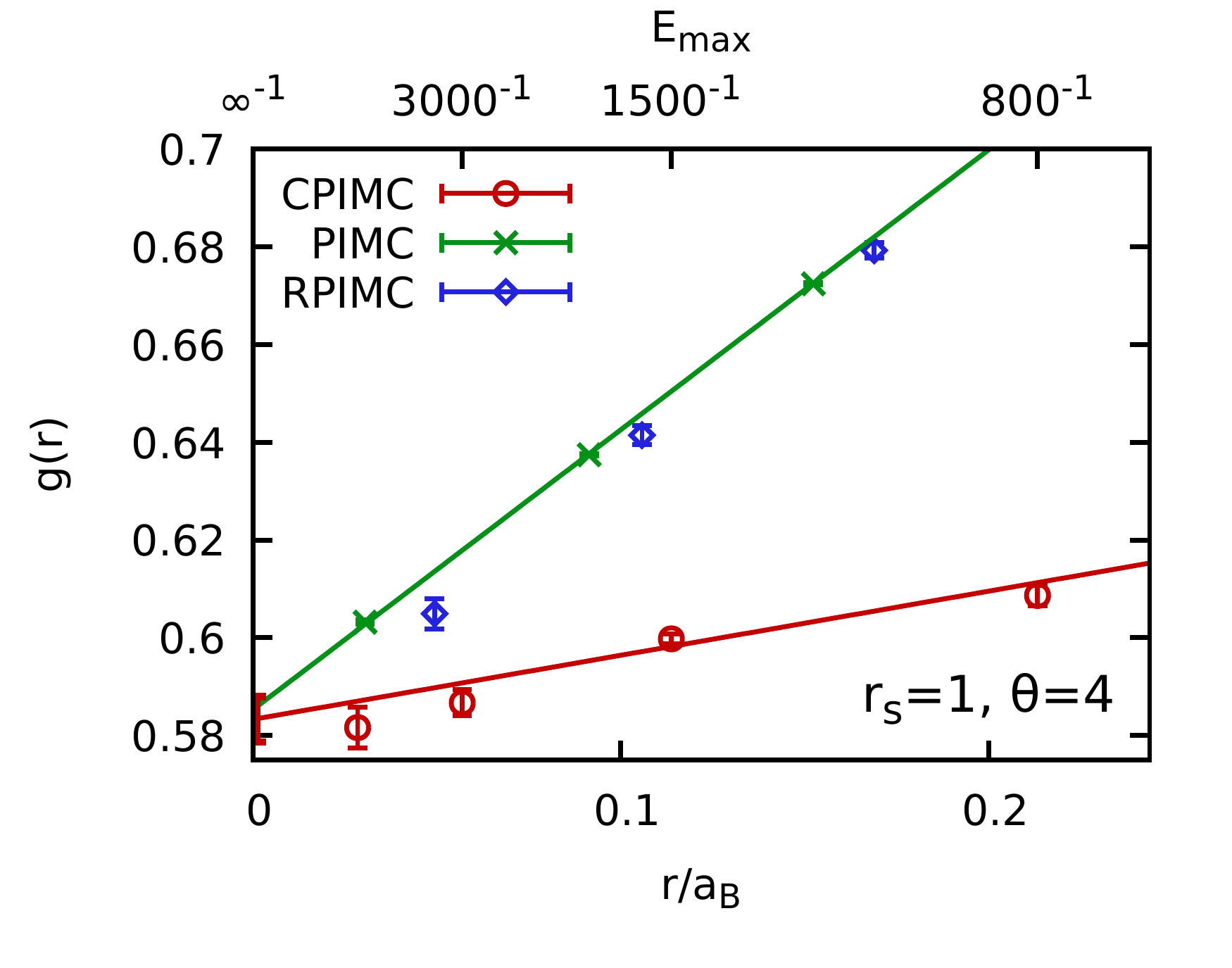}
    \caption{On-top-PDF, $g_{\uparrow\downarrow}(0)=2g(0)$, for $N=54$ electrons at $r_s=1$ and $\theta=4$, from different methods. Red circles: CPIMC results for $g_{\uparrow\downarrow}(0)$ for different values of the momentum cutoff, $E_\textnormal{max}$ (top $x$-axis); solid red line: linear fit. Green crosses: standard PIMC data for the distance-dependent PDF, $g_{\uparrow\downarrow}(r)$,  (bottom $x$-axis); solid green curve: linear fit. Blue diamonds: RPIMC data from Ref.~\cite{Brown_2014} for the same conditions, but $N=66$.
}
    \label{fig:PIMC}
\end{figure}

After analyzing CPIMC data for the large momentum tail of the distribution function we now concentrate on the coefficient in front of the asymptotic $k^{-8}$ term. According to Eq.~(\ref{eq:k8-g0}), this coefficient is entirely determined by the on-top PDF $g(0)$ which is directly accessible in quantum Monte Carlo simulations. For PIMC in coordinate space, the straightforoward way is to analyze the $r$-dependence of the PDF and subsequently extrapolate to $r=0$. Typical results are shown in Fig.~\ref{fig:PIMC} for direct fermionic (labeled ``PIMC'') and restricted (``RPIMC'') PIMC simulations. In contrast, in CPIMC a direct estimator for the on-top PDF is available, cf. 
Eq.~\ref{eq:ontop_pairdensity_cpimc_estimator_further2}, and the results are included in Fig.~\ref{fig:PIMC} with the red symbols. These results depend on the size of the single-particle basis and the corresponding cut-off energy $E_{\max}$ (top $x$-axis). Overall, for a sufficiently large basis,  very good agreement of the two independent fermionic simulations -- PIMC and CPIMC -- is observed for the parameter combinations where both are feasible.

This gives additional support for our CPIMC data, in particular for its use at low temperatures, where CPIMC provides the only \textit{ab initio} approach.
In fact, CPIMC data for $g(0)$ were already used for comparisons above. In this section we investigate the density and temperature dependence of $g(0)$. But first we explore how sensitive this value depends on the number of particles in the simulation cell.

\subsubsection{
Particle Number Dependence}\label{ss:n-dependenc}
\begin{figure}[h!]
    \centering
    \includegraphics{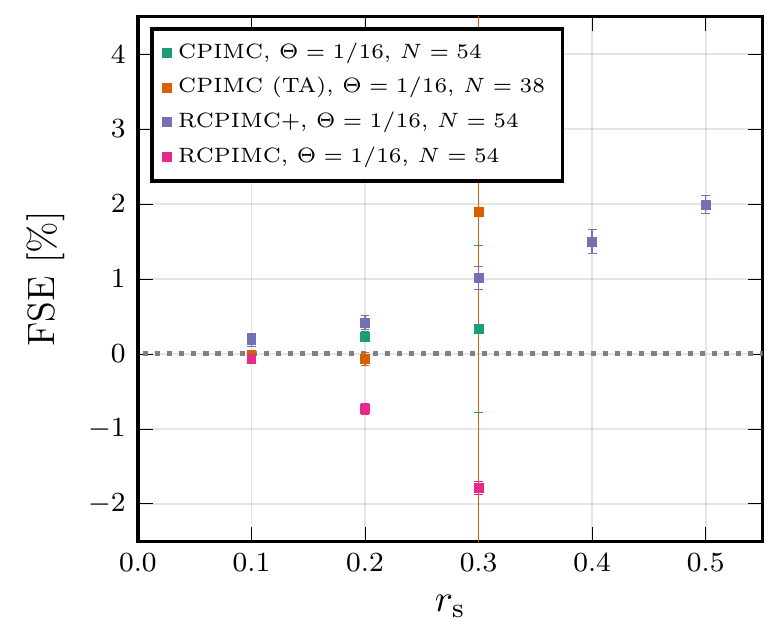}
    \caption{Influence of the particle number on the on-top PDF  at \(\Theta=0.0625\) for CPIMC  results with $N=66$, $N=54$, and $38$ particles. Shown is the relative deviation of each respective data point from the corresponding  CPIMC result for \(N=14\) which corresponds to the horizontal line at 0. Further, simulation results from the approximate RCPIMC and RCPIMC+ methods \cite{yilmaz_jcp_20} are included. ``TA'' denotes twist-angle averaging.
}
    \label{fig:gUpDown_compare_vsrs_th00625}
\end{figure}

\begin{figure}[h!]
    \centering
    \includegraphics{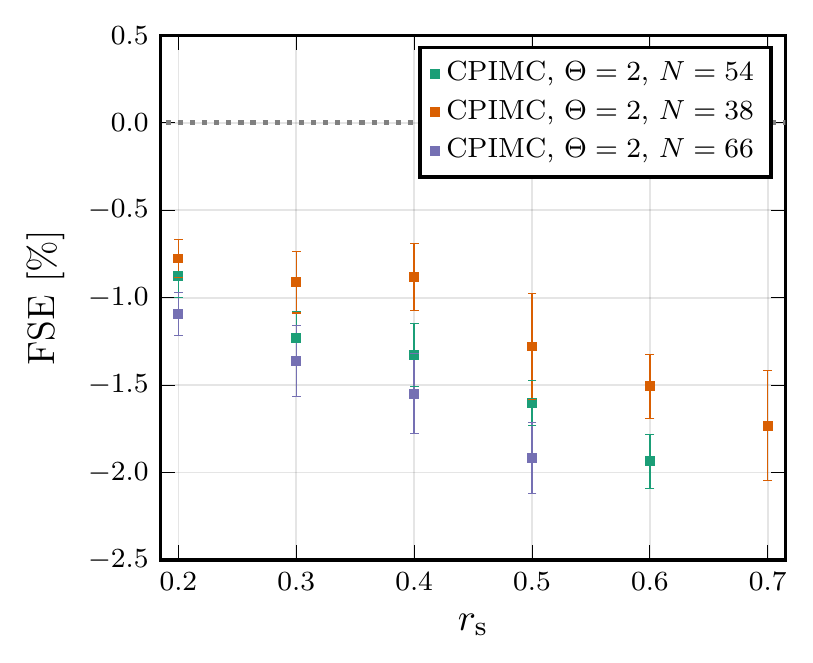}
    \caption{Same as Fig.~\ref{fig:gUpDown_compare_vsrs_th00625}, but for the  temperature $\Theta=2$.
}
    \label{fig:gUpDown_compare_vsrs_th20}
\end{figure}

We have performed extensive CPIMC simulations for $g(0)$ for a broad range of particle numbers, from $N=14$ to $N=66$. Two typical examples are shown,  for $\Theta=0.0625$, in Fig.~\ref{fig:gUpDown_compare_vsrs_th00625}, 
and for $\Theta=2$, in Fig.~\ref{fig:gUpDown_compare_vsrs_th20}. In these figures we use the case $N=14$ as the reference for comparison because, for this number, the widest range of parameters is feasible, although, naturally, simulations with larger $N$ are more accurate.
All figures confirm that finite size effects are very small in $g(0)$ and to not exceed $2\%$, even for $N=14$. 
Regarding simulations with the two approximate CPIMC variants that were discussed above \cite{yilmaz_jcp_20}, the analysis reveals that RCPIMC+ is reliable for intermediate temperatures, $0.1 \lesssim \Theta \lesssim 0.5$. Even at lower temperatures, cf. Fig.~\ref{fig:gUpDown_compare_vsrs_th00625}, we observe that RCPIMC+ data points for $N=54$ are close to CPIMC simulations for $N=54$ particles (and more accurate than CPIMC for $N=14$) and, therefore, can be well used for larger $r_s$-values, where CPIMC is not possible, due to the sign problem.  At the same time, RCPIMC \cite{yilmaz_jcp_20} turns out to be not sufficiently accurate for computing $g(0)$ and is not being used in this paper.

%
%
%
%
%

%
\subsubsection{Temperature Dependence}\label{sss:t-dependence}
\begin{figure}[h!]
    \centering
    \includegraphics{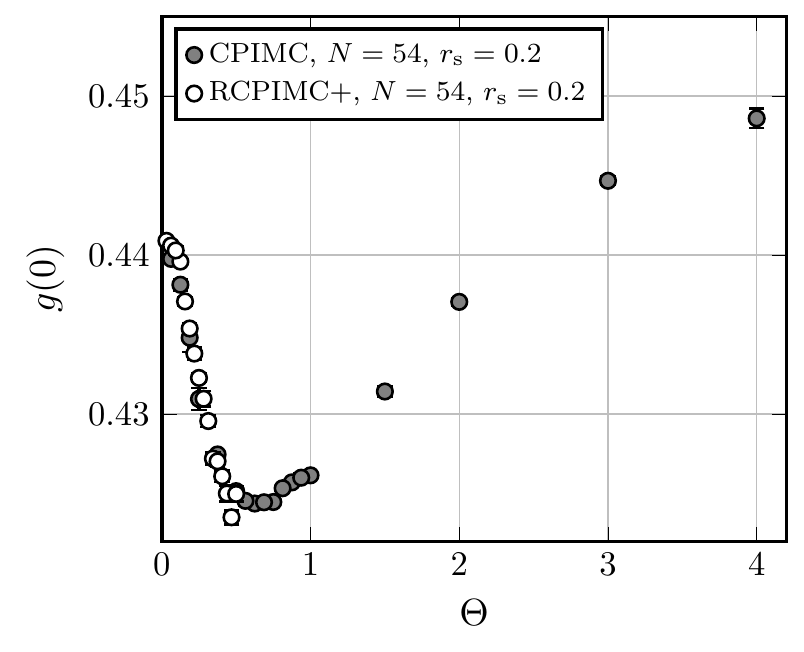}
    \caption{Temperature dependence of the on-top pair distribution for \(\rs=0.2\) from CPIMC simulations with $N=54$ particles. Very good agreement of RCPIMC+ \cite{yilmaz_jcp_20} with CPIMC is confirmed.
}
    \label{fig:g_0_vsth_rs02_N54}
\end{figure}

\begin{figure}[h!]
    \centering
    \includegraphics{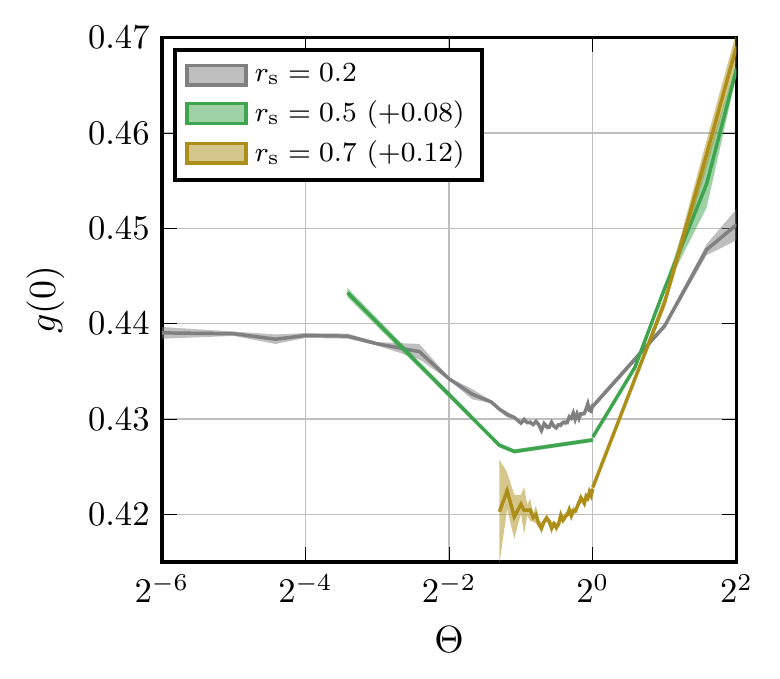}
    \caption{Temperature dependence of the on-top PDF for \(\rs=0.5\), from CPIMC simulations with 14 particles. Twist angle averaging has been applied. Shaded area indicates the statistical error. The minimum temperature is set by the fermion sign problem. For better visibility, the curves for $r_s=0.5$ and $0.7$ are shifted vertically by the number given in parantheses.
}
\label{fig:g_0_vsth_rs04_N14}
\end{figure}

We now turn to the temperature dependence of the on-top PDF. In Figs.~\ref{fig:g_0_vsth_rs02_N54} and~\ref{fig:g_0_vsth_rs04_N14}, we plot $g(0)$ from CPIMC data over a broad range of temperatures for $r_s=0.2$, and $0.2 \le r_s\le 0.7$, respectively. The figures display an interesting non-monotonic behavior: the on-top PDF increases, both towards low and high temperatures. This is easy to understand: At very low temperatures, the system approaches an almost ideal Fermi gas for which $g(0)$ would be exactly $0.5$. The (weak) Coulomb repulsion gives rise to an additional depletion of zero distance pair states. This is confirmed by the lower absolute values of $g(0)$ when $r_s$  is increased from $r_s=0.2$ to $0.4$ and $0.7$. 

On the other hand, for increasing temperature, in the range where the electron gas is dominated by classical behavior ($\Theta > 1$), both, exchange and Coulomb repulsion effects are suppressed, as compared to thermal motion, and the probability that two particles approach each other closely, tends to unity, as it would be in a non-interacting classical gas. A non-trivial question is the position of the minimum. It appears around $\Theta=k_BT/E_F \sim 0.63 $, with a depth of $0.42$, for $r_s=0.2$,  around $\Theta=0.63$, with a depth of $0.365$, for $r_s=0.4$, and around $\Theta=0.63$, with a depth of $0.296$, for $r_s=0.7$. 

This minimum can be understood as due to the balance of two opposite trends: depletion of $g(0)$, due to Coulomb repulsion and increase of $g(0)$, due to quantum delocalization effects. At high temperatures and low densities, the PDF can be expressed in binary collision (ladder) approximation 
\begin{align}
g^{\uparrow\downarrow}(r)=e^{-\beta V(r)},     
\label{eq:g-bca}
\end{align}
where $V$ is the Coulomb potential, which reproduces the behavior right of the minimum. At small interparticle distances, $r\lesssim \Lambda$, however, quantum effects have to be taken into account in the pair interaction. Averaging over the finite spatial extension of electrons leads to the replacement of the Coulomb potential by the Kelbg potential (quantum pair potential) \cite{kelbg_ap_63_1,kelbg_ap_63_2,kelbg_ap_64},
\begin{align}
    V^{\rm K}(r)=V(r)\left\{
    1-e^{-\frac{r^2}{\Lambda^2}}+\sqrt{\pi}\frac{r}{\tilde\Lambda}\left[1-{\rm erf}\left(\frac{r}{\tilde\Lambda}\right)\right]
    \right\}
    \label{eq:v-kelbg}
\end{align}
where $\tilde \Lambda=\Lambda$. Note that $V^{\rm K}$ has the asymptotic $V^{\rm K}(0;\beta)=\frac{e^2}{\Lambda(\beta)}\sim T^{1/2}$ which removes the Coulomb singularity at zero separation. While this potential has the correct derivative, $dV^{\rm K}(0)/dr=-\frac{e^2}{\Lambda^2}$, its value at $r=0$ is accurate only at weak coupling. At the same time, this potential can be extended to arbitrary coupling by retaining the same analytical form, but correcting the standard thermal DeBroglie wavelength $\Lambda$ (referring to an ideal gas) to the wave length of interacting particles, which gives rise to the so-called improved Kelbg potential \cite{filinov_jpa03, filinov_pre04},
\begin{align}
    \Lambda & \to \tilde \Lambda = \Lambda \cdot \gamma\,,\\
    V^{\rm K}(0;\beta) &\to V^{\rm IK}(0;\beta) = \frac{e^2}{\Lambda(\beta)\gamma(\beta)}\,.
\end{align}
At low temperature the effective wavelength of the electrons increases, $\gamma(\beta) \sim T^{-1/2}$, which ensures that $g^{\uparrow\downarrow}_{\rm IK}(0)=e^{-\beta V^{\rm IK}(0)}$ is finite. Accurate values for the function $\gamma$ in a two-component plasma and for different spin projections were presented in Refs.~\cite{filinov_jpa03, filinov_pre04} from a fit to PIMC data. In similar manner, the present \textit{ab initio} QMC results for the on-top PDF can be used to compute an effective DeBroglie wavelength of the warm dense uniform electron gas, and the concept of an effective quantum pair potential allows for a simple physical interpretation of some of its thermodynamic properties.

\begin{figure}[h!]
    \centering
    \includegraphics[width=\linewidth]{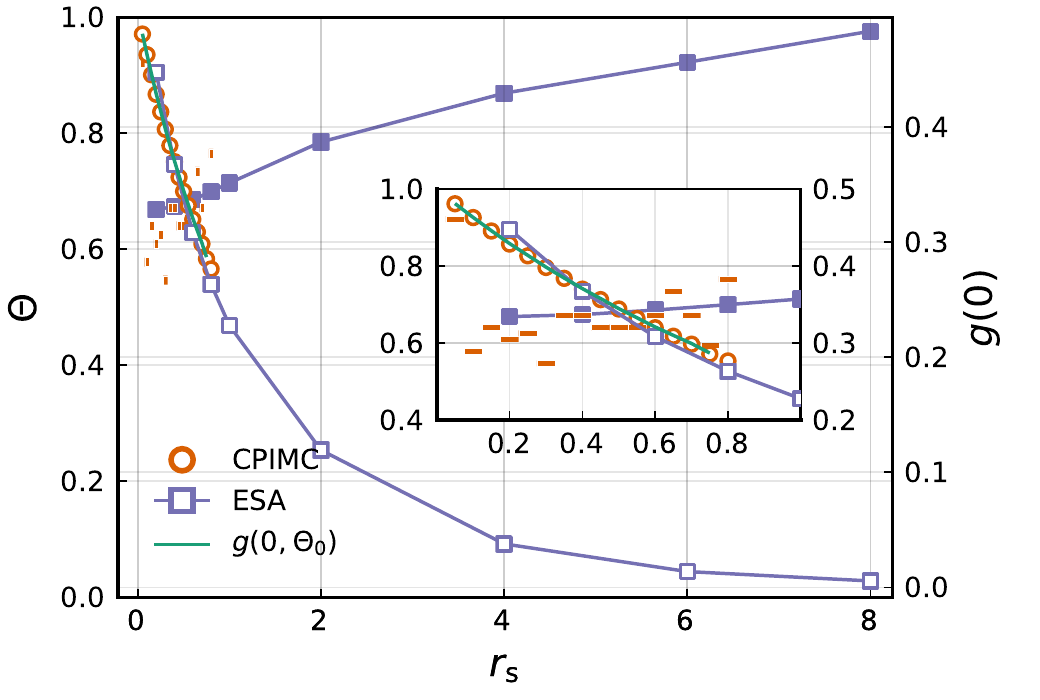}
    \caption{Analysis of the minimum of the on-top PDF. Filled symbols correspond to the location of the minimum in the \(\Theta-\rs\)--plane  (left axis \(\Theta\)). Open symbols correspond to the minimum value of the OT-PDF (right axis). Orange circles: CPIMC results for \(N=14\) particles. The green line represents the values of \(g(0)\) at a \emph{fixed} temperature \(\Theta=0.656\). ESA:  results of the extended static approximation~\cite{dornheim2020effective}, see text.
}
    \label{fig:gUpDown_minima_N14}
\end{figure}

As we already saw for the example of three densities, the location of the minimum changes with the coupling strength $r_s$. This effect is analyzed systematically in 
Fig.~\ref{fig:gUpDown_minima_N14}. We observe an increase of the minimum position, $\Theta_{\rm min}$, with $r_s$ (full squares, left axis). The reason is that, with increasing coupling, the interaction strength increases, as is seen by the increasing depth of the minimum (open symbols, right axis). Therefore, the monotonic increase of $g(r)$ with temperature sets in already at a higher temperature, when $r_s$ is increased. In addition to CPIMC data which are restricted to $r_s \lesssim 1$ we also included an analytical fit (``ESA'' \cite{dornheim2020effective}) that agrees well with CPIMC and extends the data to $r_s=8$. More information on this approximation is given in the discussion of Fig.~\ref{fig:g0comptheta1}.

\subsubsection{Density Dependence}\label{ss:density-dependencs}
\begin{figure}[h!]
    \centering
    \includegraphics{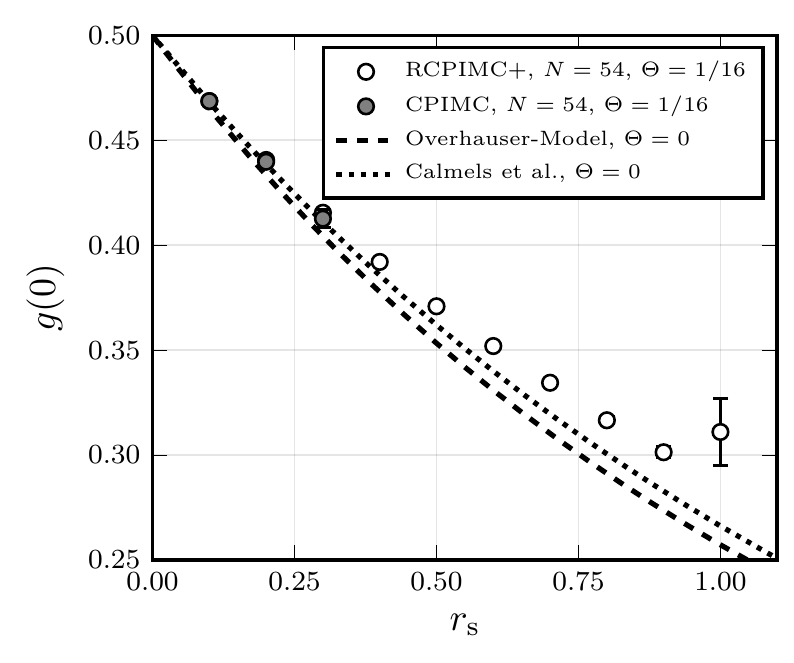}
    \caption{Density dependence of the on-top PDF for     \(\Theta=1/16\) at weak coupling. 
    Open (filled) circles: CPIMC (RCPIMC+) results for $N=54$ particles. Lines:  results of ground state models, i.e. Eq.~(\ref{eq:g0_overhausermodel}) (Overhauser model) and of Calmels \textit{et al.} Ref.~\cite{calmels_pair-correlation_1998}.}
    \label{fig:g_0_vsrs_th00625_N54}
\end{figure}

\begin{figure}[h!]
    \centering
    \includegraphics[width=0.5\textwidth]
    {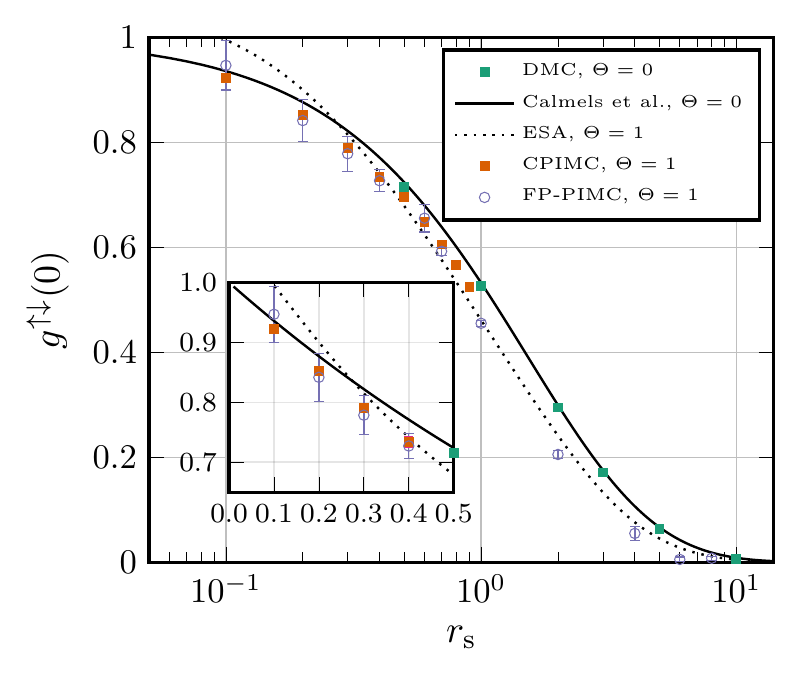}
    \caption{Density dependence of the on-top PDF for \(\Theta=1\). Red squares: CPIMC data [$r_s \leq 0.4$: $N=54$ particles; $r_s \geq 0.5$: $N=14$]. Blue circles: FP-PIMC data [$0.6 < r_s \leq 8$: $N=66$; $0.1 \leq r_s \leq 0.6$: $N=34$]. Black full (dashed) lines: ground state DMC simulations~\cite{Spink_Drummond_PRB_2013} and Eq.~(\ref{eq:g0_overhausermodel}), respectively. Inset: zoom into the high-density range (linear scale). }
    \label{fig:g0comptheta1}
\end{figure}

Let us now discuss the density dependence of the on-top PDF. As we have seen above, with increasing coupling strength, $r_s$, the value of \(\gud\)(0) decreases, due to the increased interparticle repulsion. This connection can be qualitatively understood from Eq.~(\ref{eq:g-bca})
if it is used with an effective  potential that includes many-body effects beyond the pair interaction. This monotonic decrease with $r_s$ is confirmed by our simulations for all temperatures. As an illustration, we show in Figs.~\ref{fig:g_0_vsrs_th00625_N54} and~\ref{fig:g0comptheta1} the behavior for $\Theta=0.0625$ and $\Theta=1$, respectively. 

At low temperature and weak coupling, the temperature dependence of $g(0)$ is very weak, cf. Fig.~\ref{fig:g_0_vsrs_th00625_N54}, in agreement with Fig.~\ref{fig:g_0_vsth_rs04_N14}. At $\theta=1$, finite temperature effects increase the particle repulsion due to stronger localization of electrons, and $g(0)$ falls slightly below the ground state value, cf. Fig.~\ref{fig:g0comptheta1}. This confirms the non-monotonic temperature dependence of $g(0)$ discussed above, since this temperature is in the vicinity of the minimum of $g(0)$.

\begin{figure}[h!]
    \centering
    \includegraphics[width=\linewidth]{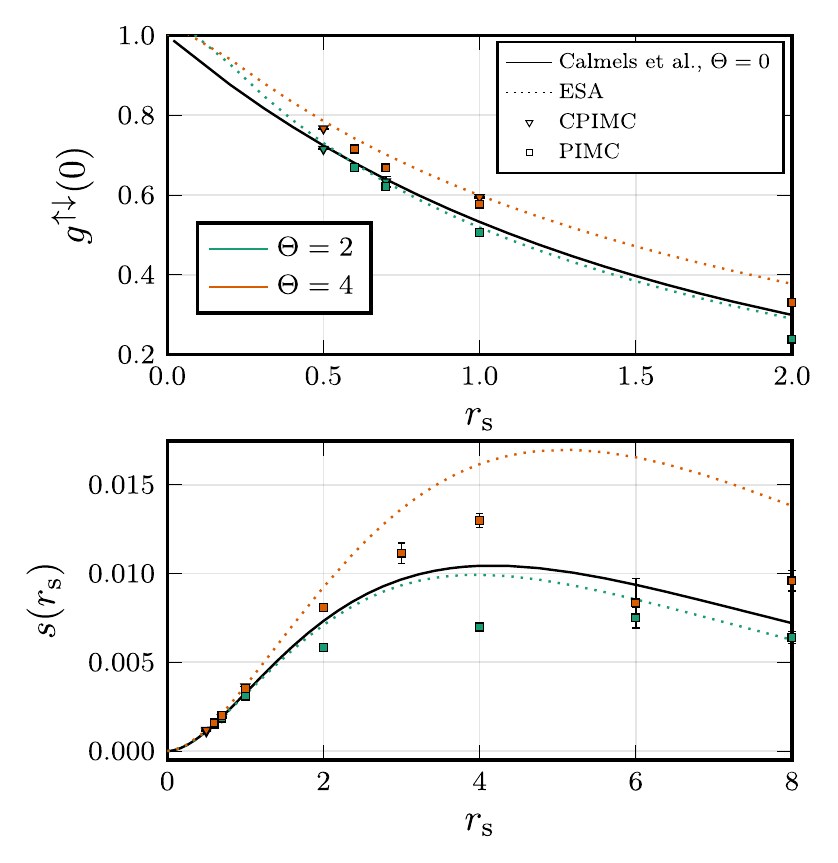}
    \caption{Top: On-top PDF (top), bottom: the function (\ref{eq:s-definition}) for two temperatures: $\Theta=4$ (orange line and symbols) and $\Theta=2$ (green line and symbols). Triangles: CPIMC data for $N=54$, squares: FP-PIMC with $N=66$ particles. Dotted lines: parametrization of Dornheim \textit{et al}.~\cite{dornheim2020effective}, black line: ground state parametrization of Calmels \textit{et al}. Note the extended $r_s$-range in the lower figure. For more data on the maximum of $s(r_s)$, see Tab.~\ref{tab:tailstrength_maxima_dornheimfit}.
}
    \label{fig:tailstrength_maximum_dornheimfit}
\end{figure}

Let us now discuss the consequences of this density and temperature dependence of $g(0)$ for the high-momentum asymptotics of $n(k)$. According to Eq.~(\ref{eq:k8-g0}), the number of electrons occupying large-$k$ states is proportional to $n(k;r_s,\Theta) \propto r^2_s\cdot g(0;r_s,\Theta)\,\frac{k^8_F(r_s)}{k^8}$, where we made the dependence on the coupling parameter explicit. Taking into account that $k_F\propto n^{1/3}\sim r_s^{-1}$, the absolute value of the asymptotic occupation number, at a given $k$ and fixed $\Theta$, scales as $n(k)\propto r_s^{-6}g(0;r_s,\Theta)\cdot k^{-8}$. On the other hand, considering the occupation number as a function of the momentum normalized to the Fermi momentum, $\kappa=k/k_F$, the density dependence becomes 
\begin{align}
n(\kappa) &\to s(r_s,\Theta)\cdot \kappa^{-8}\,,
\label{eq:n(kappa)_vs_rs}    
\\
s(r_s,\Theta) &= \frac{9}{2}\alpha^{8} r_s^2\cdot g(0;r_s,\Theta)\,,
\qquad \alpha\coloneqq \left(\frac{4}{9\pi}\right)^{\frac{1}{3}}\,.
\label{eq:s-definition}
\end{align}
Given the monotonic decrease of $g(0)$ with $r_s$, the function $n(\kappa)$ may exhibit non-monotonic behavior as a function of $r_s$, including a maximum at an intermediate $r_s$-value. This is clearly seen in Fig.~\ref{fig:tailstrength_maximum_dornheimfit} for the temperatures $\Theta=2, 4$.

As expected, at all temperatures, the coefficient $s(r_s)$ increases monotonically, for small $r_s$, starting from zero. The decrease, governed by the monotonic decrease of $g(0)$ sets in only at large $r_s$ where CPIMC simulations are not possible any more. 
On the other hand, an extensive set of restricted PIMC data~\cite{Brown_2014} for $g(r)$ is available, for $1\leq r_s\leq40$, which has recently been used by Dornheim \textit{et al.}~\cite{dornheim2020effective} to construct an analytical parametrization of $g(0;r_s,\theta)$. The results are denoted as \emph{ESA} because they constitute an important ingredient to the effective static approximation for the static local field correction that was presented in Ref.~\cite{dornheim2020effective}.

An example is shown in the lower part of Fig.~\ref{fig:tailstrength_maximum_dornheimfit} for two temperatures, $\Theta=2$ and $\Theta=4$. The maximum of $s$ is observed around $r_s=4$, for $\Theta=2$ and $r_s \approx 5$, for $\Theta=4$. 
We have performed a systematic parameter scan on the basis of the analytical fit (ESA) over a broad range of temperatures. The results are collected table \ref{tab:tailstrength_maxima_dornheimfit}. 
These results show that the maximum of $s(r_s)$ is generally located in the range $3.5 \lesssim r_s \lesssim 6.0$. Interestingly $r_s^{\rm max}$ -- the $r_s$-value where the maximum is located -- exhibits a non-monotonic temperature dependence. The reason is the non-monotonic temperature dependence of $g(0)$ that was discussed in detail in Sec.~\ref{sss:t-dependence}. Finally, the comparison with the \textit{ab initio} results contained in Fig.~\ref{fig:tailstrength_maximum_dornheimfit} suggests that the ESA fit can be further improved using our CPIMC and FP-PIMC data. 

\begin{table}[]
    \centering
    \begin{tabular}{c|c|c||c|c|c||c|c|c}
	\(\Theta\) & \(\rs^{\mathrm{max}}\) & \(s^{\mathrm{max}}\) & \(\Theta\) & \(\rs^{\mathrm{max}}\) & \(s^{\mathrm{max}}\) & \(\Theta\) & \(\rs^{\mathrm{max}}\) & \(s^{\mathrm{max}}\) \\\hline\hline
	0.0625 & 4.325 & 0.022 & 0.75 & 3.649 & 0.015  & 2.5 & 4.261 & 0.023\\\hline
	0.125 & 4.308 & 0.021 & 1.0 & 3.594 & 0.015 & 3.0 & 4.550 & 0.027\\\hline
	0.25 & 4.132 & 0.019 & 1.5 & 3.712 & 0.017 & 3.5 & 4.827 & 0.030\\\hline
    0.5	& 3.821 & 0.016 & 2.0 & 3.969 & 0.020 & 4.0 & 5.091 & 0.034\\\hline
    \end{tabular}
    \caption{Location and height of the maximum of the parameter $s(r_s)$, Eq.~(\ref{eq:s-definition}),  as a function of temperature. Results are based on the parametrization of the on-top-PDF by Dornheim \textit{et al}.~\cite{dornheim2020effective}, see also Fig.~\ref{fig:tailstrength_maximum_dornheimfit}. 
    }
    \label{tab:tailstrength_maxima_dornheimfit}
\end{table}




\subsection{Onset of the large-$k$ asymptotic of $n(k)$}\label{ss:onset}

Let us now find an approximate value of the momentum $k_\infty$ where the $k^{-8}$-asymptotic starts to dominate the behavior of the distribution function. In particular, we are interested to understand how this value
 depends on density and temperature. 
 \begin{figure}[h!]
    \centering
    \includegraphics[width=\linewidth]{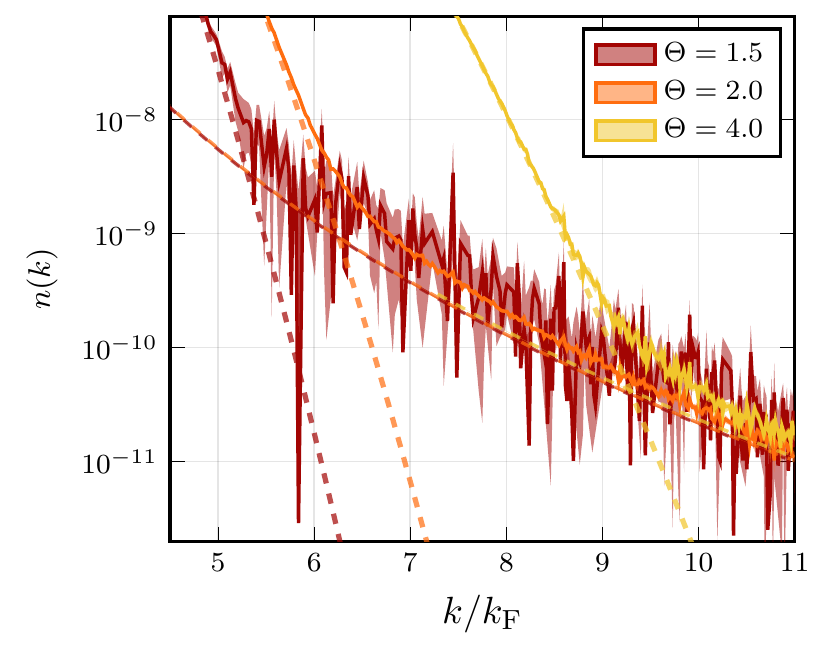}
    \caption{Illustration of the prescription (\ref{eq:onset_fermi_tail_bisect}) to determine the onset of the large-momentum asymptotic from the intersection of the ideal Fermi function, $f^{\rm id}(k)$, (dashes), with the $k^{-8}$ asymptotic, $n_\infty(k)$, (full lines of the same color). The asymptotic is determined from CPIMC simulations of the on-top PDF for $N=54$ particles. }
    \label{fig:Onset_N54}
\end{figure}
First, we observe that the significant broadening of the low-momentum part of the distribution that is observed when the temperature is increased pushes the value $k_\infty$ to larger momenta. 
Figure~\ref{fig:fullMDF_N54_rs05} suggests that this onset is near the intersection of the asymptotic, Eq.~(\ref{eq:k8-g0}), \(n^\infty(k)\) with the ideal MDF given by the Fermi-Dirac distribution function \(n^{\rm id}(k)\):
\begin{equation}
    \label{eq:onset_fermi_tail_bisect}
    n^{\rm id}(k_\infty) \overset{!}{=} n^\infty(k_\infty)\,.
\end{equation}
This approach is demonstrated in Fig.~\ref{fig:Onset_N54}, and the results are presented for a broad range of  densities, in the range of $r_s=0.2\dots 1.6$, and temperatures $\Theta \le 4$, in Fig.~\ref{fig:Onset_N14}. For this procedure, to obtain the asymptotic $n_\infty$ we used the value of \(g(0)\) that was computed in CPIMC simulations. 

\begin{figure}[h!]
    \centering
    \includegraphics[width=\linewidth]{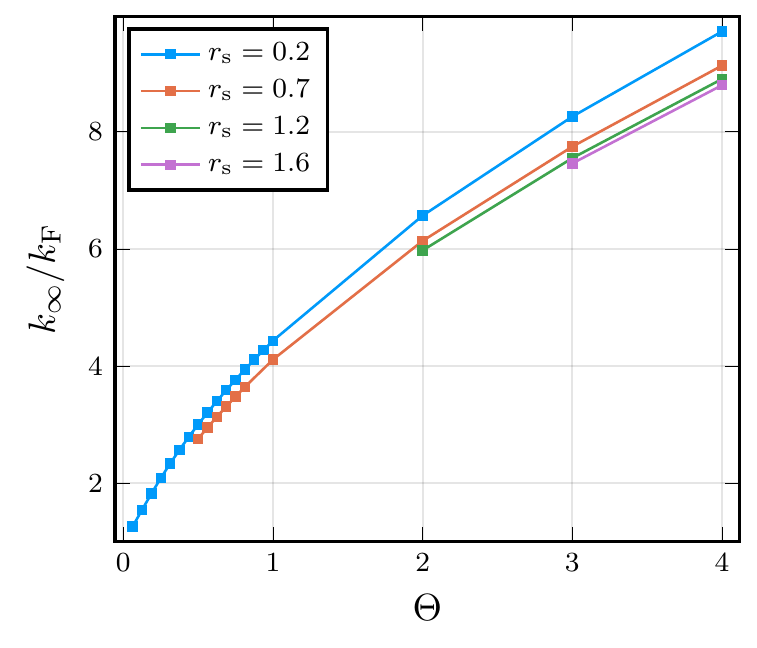}
    \caption{Onset $k_\infty$ of the large-momentum asymptotic, as calculated from Eq.~(\ref{eq:onset_fermi_tail_bisect}). 
    The procedure is illustrated in Fig.~\ref{fig:Onset_N54}. CPIMC simulations with $N=14$ particles. The lower limit of $\Theta$, for the different curves, is set by the fermion sign problem.
}
    \label{fig:Onset_N14}
\end{figure}
This figure shows that, with an increase of correlations (increase of $r_s$) the onset of the asymptotic is shifted to lower momenta, even though the dependence is weak. The figure also shows that an algebraic tail of the momentum distribution exists also in a weakly quantum degenerate plasma with $\Theta > 1$. With increasing temperature, the onset of this asymptotic is pushed to larger momenta with $k_\infty/k_F$ increasing slightly faster than $\Theta^{0.5}$.


\section{Summary and outlook} \label{s:dis}
\subsection{Summary}\label{ss:summary}
 In this paper we have performed an analysis of the momentum distribution function of the correlated warm dense electron gas using recently developed \textit{ab initio} quantum Monte Carlo methods. We have presented extensive data obtained with CPIMC, for small $r_s$. This was complemented with new fermionic propagator PIMC data, for $r_s \gtrsim 1$, so the entire density range hase been covered. Our CPIMC results for the momentum distribution of the warm uniform electron gas achieve an unprecedented accuracy -- the asymptotic is resolved up to the eleventh digit for momenta up to approximately $15 k_F$, cf. Figs.~\ref{fig:fullMDF_N54_rs05} and \ref{fig:fullMDF_N54_th2}. For all parameters the existence of the $1/k^8$ asymptotic is confirmed. Moreover, based on accurate data for the on-top PDF the absolute value of $n(k)$ in the asymptotic is obtained.
 
 While the value of the on-top PDF decreases monotonically with $r_s$, it exhibits an interesting non-monotonic temperature dependence with a minimum around $\Theta=0.656$, e.g. Fig.~\ref{fig:g_0_vsth_rs04_N14}. This was explained by a competition of Coulomb correlations and exchange effects. We also investigated the density and temperature dependence of the momentum where the algebraic decay begins to dominate the tail of the momentum distribution.
 
 In addition to the large-momentum tail we also investigated the occupation of low-momentum states in the warm dense electron gas. An interesting observation is that Coulomb interaction may lead to an enhanced occupation of low-momentum states (compared to the ideal case), which is mostly due to exchange effects, cf. Fig.~\ref{fig:Bulk_N54_th2}. Together with an enhanced population of high-momentum states this leads to a depopulation of intermediate momenta in the range $k_F \lesssim k \lesssim 3k_F$. This non-trivial re-distribution of electrons may give rise to a counter-intuitive interaction-induced decrease of the kinetic energy of the finite temperature electron gas. This confirms earlier results \cite{Militzer_PRL_2002,kraeft_pre_02} and, at the same time, complements them  extensive new and more accurate data in a broad range of parameters.

\subsection{Outlook}
Part of our results for the on-top PDF were obtained with help of the recent extended static approximation (ESA)~\cite{dornheim2020effective}. Its advantage is that it allows for relatively easy parameter scans in a broad range of densities and temperatures. Therefore, an important task is to further improve this approximation with the present high-quality data for $g(0)$.
The present simulations concentrated on the range of $r_s \lesssim 10$ which is of relevance for warm dense matter.
At the same time the jellium model is also of interest for the strongly correlated electron liquid, e.g.~\cite{dornheim_prl_20,dornheim2019strongly}. It will, therefore, be interesting to extend this analysis to larger $r_s$-values, which should be straightforward based on an analysis of the on-top PDF.

Finally, the momentum distribution function is of crucial importance for realistic two-component plasmas for which extensive restricted PIMC simulations, e.g. \cite{militzer_path_2000,hu_militzer_PhysRevLett.104.235003} and fermionic PIMC simulations, e.g. \cite{filinov_ppcf_01,bonitz_prl_5} have been performed. Therefore, an extension of the present analysis of the on-top PDF two two-component QMC simulations if of high interest.

This will also be the basis for the application of the present results to estimate the effect of power law tails in $n(k)$ in fusion rates, e.g. \cite{salpeter_69,ichimaru_RevModPhys.65.255,dewitt_ctpp.2150390124}, and other inelastic processes, that involve the impact of energetic particles. An example for the latter are electron impact excitation and ionization rates of atoms in a dense plasma. Such effects were predicted for various chemical reactions in Ref.~\cite{starostin_jetp17} based on a approximate treatment of collision rates and phenomenological Lorentzian-type broadening of the electron spectral function in Eq.~(\ref{eq:kba}). However, such approximations are known to violate energy conservation, e.g. \cite{bonitz-etal.99epjb}. The present approach to $n(k)$ makes such approximations obsolete and, moreover, eliminates the multiple integrations over the energy variables in Ref.~\cite{starostin_jetp17}, substantially simplifying the expressions for the rates. 

Finally, the relevance of algebraic tails of $n(k)$ for nuclear fusion rates in dense plasmas was discussed by many authors, e.g. \cite{savchenko_pop01, starostin_quantum_2002,starostin_jetp17,fisch_epjd12}, but the agreement with experimental data remains open. The results of the present work are applicable to many fusion reactions of fermionic particles, such as the proton-proton or $^3{\rm He}$--$^3{\rm He}$  fusion reactions in the sun or supernova stars that were considered e.g. in Refs.~\cite{ichimaru_RevModPhys.65.255,fisch_epjd12}. For quantitative comparisons the present simulations should be extended to multi-component electron-ion plasmas and include screening effects of the ion-ion interactions, e.g. \cite{dewitt_ctpp.2150390124}, which does not pose a principal problem.

 \section*{Acknowledgments}
This work has been supported by the Deutsche Forschungsgemeinschaft via project BO1366-15/1.
TD acknowledges financial support by the Center for Advanced Systems Understanding (CASUS) which is financed by the German Federal Ministry of Education and Research (BMBF) and by the Saxon Ministry for Science, Art, and Tourism (SMWK) with tax funds on the basis of the budget approved by the Saxon State Parliament.

We gratefully acknowledge CPU-time at the Norddeutscher Verbund f\"ur Hoch- und H\"ochstleistungsrechnen (HLRN) via grant shp00026 and on a Bull Cluster at the Center for Information Services and High Performance Computing (ZIH) at Technische Universit\"at Dresden.

\section*{References}
\bibliographystyle{unsrt}

\begin{appendix}
\section{Derivation of the CPIMC-Estimator for the on-top PDF, Eq.~(\ref{eq:ontop_pairdensity_cpimc_estimator_further2})}
\label{app:on-top-pdf}
We start by expressing the field operators in terms of the creation and annihilation operators in momentum representation, cf. Eqs.~(\ref{eq:creation_annihilation_operators_basis_change}),
\begin{widetext}
\begin{equation}
    \label{eq:density_correlator_1}
    \begin{aligned}
    \fcre{\sigma_1}(\vec{r}) \fcre{\sigma_2}(\vec{r}) \fan{\sigma_2}(\vec{r}) \fan{\sigma_1}(\vec{r})
    &=
    \left(\sum\limits_i \phi_i^*(\vr,\sigma_1) \cre{i} \right) 
    \left(\sum\limits_j \phi_j^*(\vr,\sigma_2) \cre{j} \right) 
    \left(\sum\limits_k \phi_k(\vr,\sigma_2) \an{k} \right)
    \left(\sum\limits_l \phi_l(\vr,\sigma_1) \an{l} \right)
    \\
    \\
    &=
    \sum\limits_{ijkl} \varphi_i^*(\vr)\varphi_j^*(\vr) \varphi_k(\vr) \varphi_l(\vr) \delta_{s_i,\sigma_1} \delta_{s_j,\sigma_2} \delta_{s_k,\sigma_2} \delta_{s_l,\sigma_1} \cre{i}\cre{j}\an{k}\an{l}
    .
    \end{aligned}
\end{equation}
%
The equation is symmetric with respect to the two possible choices of the spin projections \((\sigma_1=\uparrow,\sigma_2=\downarrow)\) and 
\((\sigma_1=\downarrow,\sigma_2=\uparrow)\), so we extend the sum over the two possibilities. Since we are interested in the case of antiparallel spins,  \(\sigma_1\neq\sigma_2\), we consider the following relations of the summation indices,

\begin{equation}
    \label{eq:spin_sum_indices_set_relation}
    \begin{gathered}
     \Set{i,j,k,l \in\zn| s_i = \sigma_1 = s_l, s_j = \sigma_2 = s_k}
    \cup
    \Set{i,j,k,l \in\zn| s_i = \sigma_2 = s_l, s_j = \sigma_1 = s_k}
    \\
    =
    \Set{i,j,k,l \in\zn| s_i = s_l, s_j = s_k}
    \setminus
    \Set{i,j,k,l \in\zn| s_i = s_l = s_j = s_k}\,.
    \end{gathered}
\end{equation}
Thus the last line of  Eq.~(\ref{eq:density_correlator_1}) can be replaced by the sum over the sets in the last line of  Eq.~(\ref{eq:spin_sum_indices_set_relation}). Since both possible choices of the spins are allowed in the latter relation, the sum is twice the value of one definite choice,

\begin{equation}
    \label{eq:density_correlator_2}
    \begin{aligned}
    \fcre{\sigma_1}(\vec{r}) \fcre{\sigma_2}(\vec{r}) \fan{\sigma_2}(\vec{r}) \fan{\sigma_1}(\vec{r})
    &=
    \sum\limits_{ijkl} \varphi_i^*(\vr)\varphi_j^*(\vr) \varphi_k(\vr) \varphi_l(\vr) \delta_{s_i,\sigma_1} \delta_{s_j,\sigma_2} \delta_{s_k,\sigma_2} \delta_{s_l,\sigma_1} \cre{i}\cre{j}\an{k}\an{l}
    \\
    &=
    \fro{2}
    \sum\limits_{ijkl} \varphi_i^*(\vr)\varphi_j^*(\vr) \varphi_k(\vr) \varphi_l(\vr) \delta_{s_i,s_l} \delta_{s_j,s_k} (1 - \delta_{s_i,s_j}) \cre{i}\cre{j}\an{k}\an{l}\,.
    \end{aligned}
\end{equation}
The statistical expectation value of this four-operator product can be expressed via the momentum representation of the two-particle density matrix, $d_{ijkl}$,
\begin{equation}
    \label{eq:density_correlator_dijkl}
    \Braket{
    \fcre{\sigma_1}(\vec{r}) \fcre{\sigma_2}(\vec{r}) \fan{\sigma_2}(\vec{r}) \fan{\sigma_1}(\vec{r})
    }
    =
    \fro{2}
    \sum\limits_{ijkl} \varphi_i^*(\vr)\varphi_j^*(\vr) \varphi_k(\vr) \varphi_l(\vr) \delta_{s_i,s_l} \delta_{s_j,s_k} (1 - \delta_{s_i,s_j})
    \underbracket{ \braket{ \cre{i}\cre{j}\an{k}\an{l} } }_{= d_{ijkl}}\,.
\end{equation}
\end{widetext}
We further need two-operator products that give rise to the spin densities appearing in the denominator of  Eq.~(\ref{eq:pair_distribution_spinresolved_def}).
Applying again the basis transformation, Eq.~(\ref{eq:creation_annihilation_operators_basis_change}), we obtain
\begin{equation}
    \begin{aligned}
    \label{eq:spin-density_operator}
    \fcre{\sigma}(\vec{r}) \fan{\sigma}(\vec{r})
    &= \sum\limits_{ij} \varphi_i(\vr) \varphi_j(\vr) \delta_{s_i,\sigma} \delta_{s_j,\sigma}\cre{i}\an{j}\,.
    \end{aligned}
\end{equation}
In the uniform electron gas, momentum conservation leads to \(\braket{\cre{i}\an{j}} = \braket{\n_i} \delta_{i,j}\) and, consequently, 

\begin{equation}
    \label{eq:spin-density}
    \begin{aligned}
    \Braket{\fcre{\sigma}(\vec{r}) \fan{\sigma}(\vec{r})}
    &= \sum\limits_{ij} \varphi_i(\vr) \varphi_j(\vr) \delta_{s_i,\sigma} 
    \delta_{s_j,\sigma}
    \Braket{\cre{i}\an{j}}
    \\
    &
    = \sum\limits_{i} \left| \varphi_i(\vr) \right|^2 \delta_{s_i,\sigma} \Braket{\n_i}\,.
    \end{aligned}
\end{equation}
The expectation value, Eq.~(\ref{eq:density_correlator_dijkl}), and the spin density, Eq.~(\ref{eq:spin-density}), contain products of plane wave single-particle orbitals~(\ref{eq:plane_waves_single_particle_basis_def}) for which 
$    \left| \varphi_i(\vr) \right|^2 = 
    \fro{V} $, and, 
 due to momentum conservation, 
\begin{align}
    \label{eq:plane_waves_pcons}
    &\varphi_i^*(\vr)\varphi_j^*(\vr) \varphi_k(\vr) \varphi_l(\vr) =
    \nonumber\\
    &\qquad\fro{V^2} \euler^{\ii\overbracket{(\vk_k + \vk_l - \vk_i - \vk_j)}^{=0}\vr} = \fro{V^2}.
\end{align}
With the definition~(\ref{eq:pair_distribution_spinresolved_def}) of the spin-resolved pair distribution function and the results from Eqs.~(\ref{eq:density_correlator_dijkl}) and~(\ref{eq:spin-density}), the on-top PDF may be expressed via  quantities that are directly accessible in CPIMC simulations,
\begin{widetext}
\begin{equation}
    \label{eq:ontop_pairdensity_cpimc}
    \begin{aligned}
    \gud_0 = g_{\sigma_1 \sigma_2}(\vr,\vr)
    &= 
    \frac{
        \fro{2V^2} \sum\limits_{ijkl} \delta_{s_i,s_l} \delta_{s_j,s_k} (1 - \delta_{s_i,s_j}) d_{ijkl}
    }{
        \left( \fro{V} \sum\limits_{i} \delta_{s_i,\sigma_1} \Braket{\n_i} \right)
        \left( \fro{V} \sum\limits_{i} \delta_{s_i,\sigma_2} \Braket{\n_i} \right)
    }
    =
    \fro{2}
    \frac{
        \sum\limits_{ijkl} \delta_{s_i,s_l} \delta_{s_j,s_k} (1 - \delta_{s_i,s_j}) d_{ijkl}
    }{
        \underbracket{\sum\limits_{i} \delta_{s_i,\sigma_1} \Braket{\n_i}}_{N_{\sigma_1}}
        \underbracket{\sum\limits_{i} \delta_{s_i,\sigma_2} \Braket{\n_i}}_{N_{\sigma_2}}
    }
    \\
    &=
    \fro{Z} \intsum\limits_C  
    \left(
    \fro{2 N_{\sigma_1}(C) N_{\sigma_2}(C)}
    \sum\limits_{ijkl} \delta_{s_i,s_l} \delta_{s_j,s_k} (1 - \delta_{s_i,s_j})
    d_{ijkl}(C) 
    \right)
    W(C)
    \end{aligned}
\end{equation}
\end{widetext}
%
The estimator can be read off the expression in the braces,
\begin{widetext}
\begin{equation}
    \label{eq:ontop_pairdensity_cpimc_estimator}
    \gud_0(C)
    =
    \fro{2 N_{\sigma_1}(C) N_{\sigma_2}(C)}
    \sum\limits_{ijkl}  \underbracket{ 
    \delta_{s_i,s_l} \delta_{s_j,s_k} (1 - \delta_{s_i,s_j})
    d_{ijkl}(C)
    }_{\coloneqq g_{ijkl}(C)}\,,
\end{equation}
where the sum can be rearranged as
\begin{equation}
    \label{eq:ontop_pairdensity_cpimc_estimator_further}
    \begin{aligned}
    \fro{2}
    \sum\limits_{ijkl}
    g_{ijkl}
    =
    \sum\limits_{\substack{k\neq i<j\neq l\\k<l}}
    \left(
    \delta_{s_i,s_l}\delta_{s_j,s_k} 
    - 
    \delta_{s_j,s_l}\delta_{s_i,s_k} 
    \right)(1 - \delta_{s_i,s_j}) d_{ijkl}
    -
    \sum\limits_{i<j}
    (1 - \delta_{s_i,s_j}) d_{ijij}
    ,
    \end{aligned}
\end{equation}
using the symmetry properties of the two-particle density matrix. The first sum is over the off-diagonal matrix elements, where the conditions of Eq.~(\ref{eq:estimator_2pdm_offdiag}) apply. The latter sum is diagonal in creation and annihilation operators, and the conditions of Eq.~(\ref{eq:estimator_2pdm_diag}) are met.
Finally, we obtain,
\begin{equation}
    \label{aeq:ontop_pairdensity_cpimc_estimator_further2}
    \begin{aligned}
    \gud_0(C)
    &=
    \fro{\beta}\sum_{\nu=1}^K 
    \sum\limits_{\substack{k\neq i<j\neq l\\k<l}}
    \left(
    \delta_{s_{j_{\nu}},s_{l_{\nu}}}\delta_{s_{i_{\nu}},s_{k_{\nu}}}
    - 
    \delta_{s_{i_{\nu}},s_{l_{\nu}}}\delta_{s_{j_{\nu}},s_{k_{\nu}}}
    \right)(1 - \delta_{s_{i{\nu}},s_{j{\nu}}}) w(\kappa_{\nu} )
    \\
    &
    -
    \sum_{\nu=0}^{K} 
    \sum\limits_{i<j}
    (1 - \delta_{s_i,s_j}) n_i^{(\nu)}  n_j^{(\nu)}\frac{\tau_{\nu+1}-\tau_\nu}{\beta}
    .
    \end{aligned}
\nonumber    
\end{equation}
\end{widetext}
The first sum extends over all kinks \(\kappa_{\nu} \coloneqq ( {i_{\nu}},{j_{\nu}},{k_{\nu}},{l_{\nu}} )\) with the proper ordering of the indices ensured by the Kronecker deltas. 
The second sum extends over all occupation numbers of occupied orbitals, \(i<j\), with opposite spin projections, at all imaginary time intervals weighted by the relative extension of the time slice in imaginary time.





\section{Modification of the kinetic energy by interaction effects}
The influence of Coulomb interaction on the kinetic energy of the warm dense uniform electron gas was studied in the main text in Sec.~\ref{sss:low-momentum-states}, see in particular Figs.~\ref{fig:Bulk_N54_rs05} and~\ref{fig:Bulk_N54_th2}. In this Appendix we provide tables with extensive benchmark data for the kinetic energy of the UEG compared to the kinetic energy of the ideal system, based on \textit{ab initio} CPIMC simulations, for temperatures $0.0625 \le \Theta \le 4$ and $r_s \le 2$. 
%
\begin{table}[]
    \centering
    \begin{tabular}{|c|c|c|c|c|}
\hline
\(N\)  & \(\Theta\)      & \(\rs\)   & \(\braket{T}^{\mathrm{id}}\) &  \(\braket{T}\)\\\hline\hline
54 & 0.0625 & 0.1 & 108.8100 &  108.8246  \(\pm\) 0.0004 \\\hline
54 & 0.0625 & 0.2 & 108.8192 &  108.9342  \(\pm\) 0.0001 \\\hline
54 & 0.0625 & 0.3 & 108.8181 &  109.1260  \(\pm\) 0.0347 \\\hline
54 & 0.125  & 0.1 & 114.4736 &  114.3502  \(\pm\) 0.0005 \\\hline
54 & 0.125  & 0.2 & 114.4478 &  114.3825  \(\pm\) 0.0109 \\\hline
54 & 0.25   & 0.1 & 136.0115 &  135.5971  \(\pm\) 0.0009 \\\hline
54 & 0.25   & 0.2 & 136.0206 &  135.4052  \(\pm\) 0.0059 \\\hline
54 & 0.5    & 0.1 & 190.1595 &  189.4961  \(\pm\) 0.0020 \\\hline
54 & 0.5    & 0.2 & 190.1765 &  189.0191  \(\pm\) 0.0018 \\\hline
54 & 0.5    & 0.3 & 190.2606 &  188.5758  \(\pm\) 0.1650 \\\hline
54 & 0.75   & 0.1 & 251.5991 &  250.7692  \(\pm\) 0.0044 \\\hline
54 & 0.75   & 0.2 & 251.5512 &  250.1747  \(\pm\) 0.0039 \\\hline
54 & 1      & 0.1 & 316.1412 &  315.5346  \(\pm\) 0.0045 \\\hline
54 & 1      & 0.2 & 316.1129 &  314.9269  \(\pm\) 0.0052 \\\hline
54 & 1      & 0.3 & 316.2673 &  314.4380  \(\pm\) 0.0069 \\\hline
54 & 1      & 0.4 & 316.2784 &  314.0231  \(\pm\) 0.0108 \\\hline
54 & 1.5    & 0.2 & 449.9120 &  448.8363  \(\pm\) 0.0116 \\\hline
54 & 1.5    & 0.4 & 449.9719 &  447.9611  \(\pm\) 0.0174 \\\hline
54 & 1.5    & 0.5 & 450.0298 &  447.5733  \(\pm\) 0.0091 \\\hline
54 & 2      & 0.2 & 586.2656 &  585.3832  \(\pm\) 0.0150 \\\hline
54 & 2      & 0.4 & 586.2975 &  584.5058  \(\pm\) 0.0227 \\\hline
54 & 2      & 0.5 & 586.5608 &  584.1818  \(\pm\) 0.0044 \\\hline
54 & 2      & 0.7 & 586.1378 &  583.5645  \(\pm\) 0.0155 \\\hline
54 & 4      & 0.2 & 1139.5097 & 1139.5308  \(\pm\) 0.0974 \\\hline
54 & 4      & 0.4 & 1139.6215 & 1138.7960  \(\pm\) 0.1286 \\\hline
54 & 4      & 0.5 & 1141.0448 & 1138.5607  \(\pm\) 0.0168 \\\hline
54 & 4      & 0.7 & 1140.7310 & 1138.0877  \(\pm\) 0.2161 \\\hline
54 & 4      & 1   & 1139.7371 & 1137.4673  \(\pm\) 0.0257 \\\hline
54 & 4      & 1.3 & 1140.4339 & 1137.4851  \(\pm\) 0.4838 \\\hline
\end{tabular}
    \caption{Comparison of the kinetic energy of an ideal system with the one of an interacting system. CPIMC results using 54 particles. The errors in the 4th column are lower than \(10^{-5}\).}
    \label{tab:TvsT0_table_N54}
\end{table}


\begin{table}[]
    \centering
    \begin{tabular}{|c|c|c|c|c|}
\hline
\(N\)  & \(\Theta\)      & \(\rs\)   & \(\braket{T}^{\mathrm{id}}\) &  \(\braket{T}\)\\\hline\hline
14 & 0.0625 & 0.1 & 12.0017 &   12.0066  \(\pm\) 0.0    \\\hline
14 & 0.0625 & 0.2 & 12.0014 &   12.0202  \(\pm\) 0.0    \\\hline
14 & 0.0625 & 0.3 & 12.0011 &   12.0412  \(\pm\) 0.0    \\\hline
14 & 0.0625 & 0.4 & 12.0015 &   12.0685  \(\pm\) 0.0    \\\hline
14 & 0.0625 & 0.5 & 12.0015 &   12.1010  \(\pm\) 0.0    \\\hline
14 & 0.125  & 0.1 & 12.4235 &   12.4326  \(\pm\) 0.0    \\\hline
14 & 0.125  & 0.2 & 12.4181 &   12.4439  \(\pm\) 0.0    \\\hline
14 & 0.125  & 0.3 & 12.4191 &   12.4627  \(\pm\) 0.0    \\\hline
14 & 0.125  & 0.4 & 12.4308 &   12.4885  \(\pm\) 0.0    \\\hline
14 & 0.125  & 0.5 & 12.4429 &   12.5193  \(\pm\) 0.0    \\\hline
14 & 0.25   & 0.2 & 15.1806 &   15.1710  \(\pm\) 0.0    \\\hline
14 & 0.25   & 0.3 & 15.1742 &   15.1796  \(\pm\) 0.0    \\\hline
14 & 0.25   & 0.4 & 15.1805 &   15.1965  \(\pm\) 0.0    \\\hline
14 & 0.25   & 0.5 & 15.1856 &   15.2205  \(\pm\) 0.0001 \\\hline
14 & 0.25   & 0.6 & 15.1954 &   15.2499  \(\pm\) 0.0011 \\\hline
14 & 0.5    & 0.1 & 20.1999 &   20.1744  \(\pm\) 0.0001 \\\hline
14 & 0.5    & 0.2 & 20.1995 &   20.1509  \(\pm\) 0.0003 \\\hline
14 & 0.5    & 0.3 & 20.2131 &   20.1370  \(\pm\) 0.0002 \\\hline
14 & 0.5    & 0.4 & 20.2276 &   20.1307  \(\pm\) 0.0001 \\\hline
14 & 0.5    & 0.5 & 20.2184 &   20.1321  \(\pm\) 0.0001 \\\hline
14 & 0.5    & 0.6 & 20.1993 &   20.1395  \(\pm\) 0.0001 \\\hline
14 & 0.5    & 0.7 & 20.1976 &   20.1522  \(\pm\) 0.0010 \\\hline
14 & 0.75   & 0.2 & 26.1944 &   26.1617  \(\pm\) 0.0002 \\\hline
14 & 0.75   & 0.3 & 26.2597 &   26.1323  \(\pm\) 0.0003 \\\hline
14 & 0.75   & 0.4 & 26.2478 &   26.1094  \(\pm\) 0.0003 \\\hline
14 & 0.75   & 0.5 & 26.2289 &   26.0942  \(\pm\) 0.0003 \\\hline
14 & 0.75   & 0.6 & 26.2855 &   26.0838  \(\pm\) 0.0003 \\\hline
14 & 0.75   & 0.7 & 26.2125 &   26.0797  \(\pm\) 0.0003 \\\hline
14 & 0.75   & 0.8 & 26.2691 &   26.0798  \(\pm\) 0.0009 \\\hline
14 & 1      & 0.1 & 33.0213 &   32.9558  \(\pm\) 0.0006 \\\hline
14 & 1      & 0.2 & 33.0236 &   32.9127  \(\pm\) 0.0006 \\\hline
14 & 1      & 0.3 & 33.0191 &   32.8765  \(\pm\) 0.0004 \\\hline
14 & 1      & 0.4 & 32.9910 &   32.8460  \(\pm\) 0.0004 \\\hline
14 & 1      & 0.5 & 33.0087 &   32.8217  \(\pm\) 0.0003 \\\hline
14 & 1      & 0.6 & 33.0067 &   32.8029  \(\pm\) 0.0003 \\\hline
14 & 1      & 0.7 & 32.9989 &   32.7882  \(\pm\) 0.0001 \\\hline
14 & 1      & 0.8 & 32.9953 &   32.7776  \(\pm\) 0.0003 \\\hline
14 & 1      & 0.9 & 32.9874  &   32.7713  \(\pm\) 0.0010 \\\hline
14 & 1      & 1   & 33.0208 &   32.7629  \(\pm\) 0.0158 \\\hline
14 & 1.5    & 0.1 & 47.1352 &   47.0962  \(\pm\) 0.0004 \\\hline
14 & 2      & 0.1 & 61.5714 &   61.5066  \(\pm\) 0.0009 \\\hline
14 & 2      & 0.2 & 61.5276 &   61.4679  \(\pm\) 0.0007 \\\hline
14 & 2      & 0.4 & 61.5819 &   61.3997  \(\pm\) 0.0011 \\\hline
14 & 2      & 0.5 & 61.6823 &   61.3725  \(\pm\) 0.0014 \\\hline
14 & 2      & 0.7 & 61.5246 &   61.3244  \(\pm\) 0.0016 \\\hline
14 & 2      & 1   & 61.6415 &   61.2688  \(\pm\) 0.0020 \\\hline
14 & 4      & 0.1 & 119.7877 &  119.9513  \(\pm\) 0.0018 \\\hline
14 & 4      & 0.2 & 119.9672 &  119.9236  \(\pm\) 0.0032 \\\hline
14 & 4      & 0.4 & 119.8911 &  119.8764  \(\pm\) 0.0035 \\\hline
14 & 4      & 0.5 & 119.8763 &  119.8596  \(\pm\) 0.0036 \\\hline
14 & 4      & 0.7 & 120.2538 &  119.8108  \(\pm\) 0.0043 \\\hline
14 & 4      & 1   & 119.9191 &  119.7682  \(\pm\) 0.0055 \\\hline
14 & 4      & 2   & 119.9626 &  119.6560  \(\pm\) 0.0067 \\\hline
\end{tabular}
    \caption{Same as table~\ref{tab:TvsT0_table_N54}, but for 14 particles.}
    \label{tab:TvsT0_table_N14}
\end{table}

\end{appendix}
\end{document}